\documentclass[12pt]{amsart}

\usepackage{amsmath}
\usepackage{amssymb}
\usepackage{amsthm}

\usepackage{slashed}
\usepackage{color}
\usepackage{esint}

\usepackage{graphicx}
\usepackage[all,cmtip]{xy}
\usepackage{hyperref}

\newtheorem{rmk}{Remark}

\hypersetup{
    pdfborder = {0 0 0},%
    colorlinks,%
    citecolor=blue,%
    filecolor=black,%
    linkcolor=red,%
    urlcolor=green
}

\def\k{\mathbf{k}}
\def\T{\mathcal{T}}
\def\Z{\mathbb{Z}}

\def\B{\mathcal{B}}
\def\V{\mathcal{V}}
\def\L{\mathcal{L}}
\def\R{\mathbb{R}}
\def\C{\mathbb{C}}
\def\H{\mathbb{H}}
\def\Hilb{\mathcal{H}}
\def\Fix{\boldsymbol{\Gamma}}
\def\bk{\mathbf{k}}
\def\TT{\mathbb{T}}
\def\pf{\mathrm{Pf}}
\def\CC{\mathfrak {H}_{Kr}}
\def\CQ{\mathfrak{H}^{\Theta}}
\def\Bloch{\B}
\def\BlochT{(\Bloch,\Theta)}
\def\realdet{Det_{\Theta}\Bloch_{\boldsymbol{\Gamma}}}

\begin{document}

\title{  Notes on topological insulators }
\author{Ralph M. Kaufmann} \address{Department of Mathematics, Purdue University,
		150 N. University St, West Lafayette, IN 47907}\email{rkaufman@math.purdue.edu  }
\author{Dan Li}	\address{Department of Mathematics, Purdue University,
		150 N. University St, West Lafayette, IN 47907}\email{li1863@purdue.edu   }	
        \author
{Birgit  Wehefritz-Kaufmann} \address{Department of Mathematics and Department of Physics and Astronomy, Purdue University,
		150 N. University St, West Lafayette, IN 47907}  
		\email {ebkaufma@math.purdue.edu}
 
	\keywords{Topological insulators, index theory, topological K-theory}

\date{}

\begin{abstract}
This paper is a survey of the $\mathbb{Z}_2$-valued invariant of topological insulators used in condensed matter physics.
The $\mathbb{Z}$-valued topological invariant, which was originally called the TKNN invariant in physics,  has now been fully understood as the first Chern number.
The $\Z_2$ invariant is more mysterious, we will explain
its equivalent descriptions  from different points of view and provide the relations between them.
These invariants provide the classification of topological insulators with different symmetries in which K-theory plays an important role.
Moreover, we establish that both invariants are realizations of index theorems which can also be understood in terms of condensed matter physics.

\end{abstract}
\maketitle
%\newpage

\tableofcontents

%\newpage

\section{Introduction}\label{Intro}

Topological insulators \cite{HK10} are new materials observed in nature which behave like insulators in the bulk
but have conducting edge or surface states on the boundary.
In physical terms, topological insulators are gapped electronic systems which show topologically protected non-trivial phases in the presence of  the time reversal $\mathbb{Z}_2$-symmetry. 
Because of the (odd) time reversal symmetry, topological insulators are characterized by a $\mathbb{Z}_2$-valued invariant. We will  use
$\mathbb{Z}_2$  to denote the group $\mathbb{Z}/2\mathbb{Z}$, since this is the notation prevalent  in the physics literature. In the framework of symmetry protected topological (SPT) orders \cite{GW09},
the (integer) quantum Hall effect is also viewed as a topological insulator in a broad sense, which is protected by a $U(1)$-symmetry and characterized by a $\mathbb{Z}$-valued invariant.

In this paper, we review the topological $\mathbb{Z}/\mathbb{Z}_2$ invariants that have appeared
in 2d and 3d topological insulators and show that they are realistic applications of index theory and K-theory in condensed matter physics. 
We stress that the $\mathbb{Z}_2$ index of time reversal invariant topological insulators can be understood as a mod 2 index theorem. This allows us to link together several different representations
and thus explain them as instances of a common phenomenon.

For example, the Atiyah--Singer index theorem \cite{AS68} can be used to compute the analytical index of an elliptic operator
by its topological index. In physics, studying the spectral flow of observables, such as counting the net change of quantum charges, can be converted to computing
topological invariants, e.g.\ evaluating the Chern character in gauge theory. This evaluation can then be seen as an application of the index theorem, which of course has intimate relation with K-theory.

The integer quantum Hall effect gives a first example of a topological insulator.
The first Chern number derived from the underlying geometry, which used to be called the TKNN invariant \cite{TKNN82},
describes the quantized Hall conductivity and
characterizes the integer quantum Hall effect. More precisely,  the Hall conductance can be calculated by
 the integral of a first Chern class over the Brillouin torus $\mathbb{T}^2$,
which is often compared to the Gauss--Bonnet theorem on surfaces \cite{AOS03}. The
quantum Hall effect will be discussed from a topological point of view in section \ref{QHE}.

Its cousin, the quantum spin Hall effect \cite{MHZ11} was first studied in graphene, and later observed in HgTe quantum wells
with stronger spin-orbit coupling \cite{K07}.
More recently, 3d  topological insulators were also observed in nature \cite{HK10}.
These 2d and 3d topological insulators are characterized by a $\mathbb{Z}_2$-valued  invariant, and they are
stable under small perturbations by impurities or defects, so such materials are called $\mathbb{Z}_2$ topological insulators. Physically, the $\mathbb{Z}_2$ invariant counts the parity of Majorana zero modes, 
which are quasi-particles slightly different from Majorana spinors.

Since there are many equivalent definitions of the $\mathbb{Z}_2$ invariant, one main goal of this survey is to show that
 different  descriptions of the $\mathbb{Z}_2$ invariant can be unified by a mod 2 index first introduced
 by Atiyah and Singer in \cite{AS71}.
The main fact is that taking into account the time reversal $\mathbb{Z}_2$ symmetry changes the  geometry and topology of the Brillouin torus. This  is the reason for the appearance of 
both the topological  $\mathbb{Z}_2$ invariant and the mod 2 index theorem.

Physicists originally proposed two $\mathbb{Z}_2$ invariants from different perspectives.
The Kane--Mele invariant  was first introduced in the study of the quantum spin Hall effect on graphene \cite{KM05},
and generalized to 3d  topological
insulators later \cite{FKM07,MB07}. The Kane--Mele invariant  is defined as the product of  signs of Pfaffians over the fixed points of the time reversal symmetry.
As an alternative to the discrete Pfaffian formalism of the Kane--Mele invariant, the Chern--Simons invariant has an integral form
defined by the variation of Chern--Simons actions under a specific gauge transformation induced by the time reversal symmetry \cite{QHZ08}.
In the language of index theory, the Kane--Mele invariant is an analytical index and the Chern--Simons invariant is a topological index.
Together they fit perfectly into a mod 2  index theorem. So we view them as different aspects of the same $\mathbb{Z}_2$ invariant,
and we call both of them the topological $\mathbb{Z}_2$ invariant from now on.
In section \ref{Z2Inv}, we will look into the topological $\mathbb{Z}_2$ invariant and its variants from different points of view. We then show how these definitions can be viewed 
as a mod 2 analytical or topological index in section \ref{indexpar}.

In order to model the band structure of a topological insulator, we consider  the associated Hilbert bundle, usually called the Bloch bundle,
over the Brillouin torus.
When an additional time reversal symmetry is present, the Bloch bundle becomes a so-called Quaternionic vector bundle, see \S\ref{bundletypepar}.
 This is the reason that  Quaternionic K-theory $KH$ (or $KQ$),  and  with it ---via isomorphism theorems--- Real K-theory $KR$,  plays an important role in the classification
problem of topological insulators. More generally, when one considers general finite symmetry groups in condensed matter physics,
twisted equivariant K-theory can be used to classify
topological phases with symmetries \cite{FM13}. A brief discussion about the K-theoretic classification of topological phases
is also included in section \ref{KThry}.
We conclude with further discussions of an index theorem realizing the bulk--edge correspondence and possible generalizations to noncommutative geometry.

In this survey, we collect ideas and results from both the physical and mathematical literature in order to link them ---by exhibiting the mechanisms under which they are equivalent---  and put them into perspective under the common aspect
of index theorems.
We focus on those creative ideas and write down the most careful treatment to date of those constructions, but at each step provide the necessary setup and references where the relevant details can be found. 
We also include some original ideas of our own, for example, we derive a formula of the $\mathbb{Z}_2$ invariant based on holonomy in \S\ref{holpar}. This survey is different from others 
since we put the topological $\mathbb{Z}/\mathbb{Z}_2$ invariants in the framework of index theory and K-theory.

As the literature of topological insulators is already very vast, we apologize in advance if we inadvertently missed some material.

\section{First Chern number as the $\Z$ invariant} \label{QHE}
In this section, we will review the integer quantum Hall effect (QHE), the first Chern number, and the bulk-boundary correspondence of QHE as an index theorem.
The purpose of this section is twofold.
First,
the quantum Hall effect provides the prototype of a topological insulator, in which the only topological invariant is given by the first Chern number. 
It is helpful to understand the geometric picture of this simplest 2d model before further generalizations.
The second purpose is to familiarize the  reader
 with the basic language used in condensed matter physics.

In the classical Hall system, a magnetic field ${B}$ is turned on perpendicular to a plane of finite size. Then one
applies an electric current and
considers a single electron flowing in  one direction in the plane, say the y-direction.
The Hall conductance is computed as the ratio of
the current density to  the electric field,
\begin{equation*}
 \sigma_{xy} = \frac{j_y}{E_x} ={\nu}\frac{e^2}{h}
\end{equation*}
where $\nu$ is  the so--called filling factor,
$e$ is the elementary charge and $h$ is the Planck constant.

When the magnetic field is very strong and the temperature is low enough, the filling factor is quantized and takes on  positive integer values. This phenomenon is called the integer quantum Hall effect,
\begin{equation}
  \sigma_{xy} = n \frac{e^2}{h}, \quad n \in \mathbb{N}
\end{equation}
The quantum number $n$, also called the TKNN integer,  is successfully understood as the first Chern number
of the Berry curvature over the Brillouin torus.

For an even stronger magnetic field and at extremely low temperature, the fractional quantum Hall effect can be observed, in which
the filling factor can also take on fractional values.
The fractional QHE will not be
discussed  here, since we will only consider Hamiltonians
with weak interactions in this survey while in the fractional case the Coulomb interaction between electrons is dominant.

\subsection{Landau levels}
\label{landauedgepar}
As a warm up,
we consider the spectral theory of a free Hamiltonian in the Landau gauge in this subsection.  Its spectrum consists of the Landau levels,
which is one of the simplest
and most important band structures in condensed matter physics.

Let us first forget about the spin property of  electrons  for the moment.  A free electron in a magnetic field has the Hamiltonian
\begin{equation*}
   H = \frac{1}{2m} (\mathbf{p} +e \mathbf{A})^2
\end{equation*}
Since the magnetic field $\mathbf{B}$ is assumed to be perpendicular to the $xy$-plane, we consider the Landau gauge $\mathbf{A} = (0, Bx, 0)$ such that
$\mathbf{B} = \nabla \times \mathbf{A} = (0, 0, B)$, in this gauge
\begin{equation*}
   H =\frac{1}{2m} \left[p_x^2 + (p_y + eBx)^2\right]
\end{equation*}
It is obvious that the commutator $[H, p_y] = 0$, so $H$ and $p_y$ can be diagonalized simultaneously.
In other words, in the Hamiltonian $p_y$ may be replaced by its eigenvalue $k_y$,
\begin{equation*}
    H =\frac{1}{2m} \left[p_x^2 + (k_y + eBx)^2\right]
\end{equation*}
Using the so--called cyclotron frequency $\omega = {eB}/{m}$, the Hamiltonian is written as
\begin{equation}
\label{edgehameq}
   H= \frac{p_x^2 }{2m}+ \frac{1}{2} m \omega^2(x + \frac{k_y}{m\omega})^2
\end{equation}
which takes the same form as the quantum harmonic oscillator.
Therefore, the energy spectrum  is given by
\begin{equation*}
   E_n = \left( \frac{1}{2} + n \right) \hbar \omega, \quad n\geq 0
\end{equation*}
 The set of wave functions (eigenspace) with energy corresponding to an integer $n$ is called the $n$-th Landau level. Denoting the harmonic oscillator eigenstates by $\psi_n$, the wave function of the free electron in our model then is
expressed as
\begin{equation} \label{edgestate}
   \phi_n(x, k_y) = e^{ik_y y}\psi_n(x + x_0), \quad x_0 = \frac{k_y}{m\omega}
\end{equation}
This state \eqref{edgestate} is the first example of an edge state,
which  by definition is the eigenfunction satisfying the eigenvalue equation of the edge Hamiltonian (\ref{edgehameq}).

In general, given a bulk Hamiltonian $H_b(x,y)$ which has translational invariance along one direction ---say the y-direction--- an edge Hamiltonian is obtained from the bulk Hamiltonian by a partial Fourier transform. 
The momenta  $k_y$ in $y$ direction
can then be treated as a parameter, so that the edge Hamiltonian becomes a one--parameter family $H_e(x,k_y)$ as do the energy levels and the eigenfunctions. I.e.\ we have the following family of equations:
\begin{equation*}
   H_e(x,k_y)\phi_n(x,k_y) = \varepsilon_n(k_y) \phi_n(x,k_y)
\end{equation*}
parameterized by  $k_y$.

In topological band theory, the band structure, i.e., the Landau levels, defines a complex vector bundle over the momentum space, more details will be given in the next subsection. 
In the simplest case, the band structure of a free Hamiltonian gives rise to a trivial vector bundle.

\subsection{Bloch bundle}
Thouless, Kohmoto, Nightingale and den Nijs (TKNN) \cite{TKNN82} first showed that
 the quantum number $n$ in the Hall conductance $\sigma_{xy} = n {e^2}/{h}$, which is called the TKNN invariant by physicists,
can be topologically identified with the first Chern number.
In this subsection, we will review the first Chern number of the quantum Hall effect based on topological band theory.

For the 2d square lattice $\Gamma \simeq \mathbb{Z}^2 \subset \mathbb{R}^2$, its Pontryagin dual is the 2-torus $\mathbb{T}^2$, and
it will be called the Brillouin torus from now on.
In condensed matter
physics,  a space dual to the original lattice is called the Brillouin zone $B$. In physics parlance,
this is the momentum space in which the crystal wave vector $\mathbf{k}$ lives.

For electrons in a crystal, the electronic states are  described by Bloch wave functions
\begin{equation}
    \psi(\mathbf{r})  = e^{i \mathbf{k} \cdot \mathbf{r}} u(\mathbf{r}),
    \quad  u(\mathbf{r} +\mathbf{R}) = u(\mathbf{r}), ~~\forall ~\mathbf{R} \in \Gamma
\end{equation}
In the following, we use the momentum representation.
The quasi--periodic eigenstates $u_n(\mathbf{k})$  are by definition solutions to
the eigenvalue equation,
\begin{equation*}
   H(\mathbf{k}) u_{n} (\mathbf{k}) = E_n(\mathbf{k}) u_{n} (\mathbf{k})\,
\end{equation*}
where $E_n(\mathbf{k})$ is called the $n$-th band function.
We assume that the  $E_n(\mathbf{k})$ are non-degenerate everywhere (i.e., no band crossing). In general, the energy spectrum $\{ E_n(\mathbf{k}), n \geq 1 \}$ is referred to as the  band structure of the model.

 Mathematically, the $u_{n}( \mathbf{k}) $ are sections of a Hilbert bundle $E\to B$ whose fiber over $\mathbf{k}\in B$ is
given by the $u_{n}( \mathbf{k})$ with fixed $\mathbf{k}$. Here one uses $L^2$ sections and the compact--open topology; see e.g.\ \cite{FM13} for a detailed study.
 All the sections  form a Hilbert space, which will be called $\mathcal{H}$. The sections   $u_{n}( \mathbf{k})$ with fixed $n$ form the sub--Hilbert space $\mathcal{H}_n$ and $\mathcal{H}=\bigoplus_n\mathcal{H}_n$.

 In the physical literature sections are called  physical states  and are written as $ |u_{n}( \mathbf{k}) \rangle$, or $ | n, \mathbf{k}\rangle$ to emphasize the quantum numbers. 
 Furthermore they are always assumed to be normalized so that $\langle n, \mathbf{k} | n, \mathbf{k} \rangle =1 $.

There are two types of bands: the occupied bands, literally occupied by the electrons, and the empty bands.
The valence band is the top one among the occupied bands and the conduction band is the bottom one among the empty bands.
The band gap is defined as the energy gap between the valence and conduction bands.
The Fermi energy $E_F$ is defined such that in the ground state  all single-particle levels with smaller energy are occupied and there is no occupied level above this energy. 
Loosely, it is the top of the collection of occupied electron energy levels. It coincides with the chemical potential at absolute zero temperature \cite{AM76}, which is the most precise definition. 
What the position of this energy inside the band structure is, depends on the specific material. For insulators $E_F$ lies in a band gap, which we will now assume.

The construction of the bundles is valid for generalized Brillouin zones such as $S^d$ or $\mathbb{T}^d$,
but in the rest of this subsection we only use  $\mathbb{T}^2$ for concreteness, which is the Brillouin zone in the 2d square lattice quantum Hall model.
Let us consider the  rank one (aka.\ line) bundle for the $n$-th occupied band over the Brillouin torus.
\begin{equation}
   \L_n \rightarrow \mathbb{T}^2,
\end{equation}
This is the eigenbundle to the $n$--th energy band $E_n$, which is assumed to be below the Fermi energy. Notice that since the eigenvalues of the Hamiltonians are real, there is an order on them. 
Its sections are the $n$-th eigenstates $|u_{n}(\mathbf{k})  \rangle: \mathbb{T}^2 \rightarrow \L_n; \mathbf{k} \mapsto   u_{n}(\mathbf{k}) $.
Assume that there exist $N$ occupied bands, then the Bloch bundle is defined  to be the  rank N bundle given by the Whitney sum,
\begin{equation*}
   \Bloch = \oplus_{n=1}^N \L_n \rightarrow \mathbb{T}^2
\end{equation*}

The Bloch bundle is a finite rank Hilbert bundle, whose space of sections is the physical Hilbert space. The sections themselves are Bloch wave functions.

The Berry connection is then defined by a covariant derivative on the Bloch bundle \cite{FM13}, whose local form on the component $\L_n$ is given by
\begin{equation}
    a^n(\mathbf{k}) = i \langle {n, \mathbf{k}}| d |{n, \mathbf{k}} \rangle
\end{equation}
so the Berry curvature on  $\L_n$ has a local expression
\begin{equation}
   f^n (\mathbf{k}) = i
   \; d \langle {n, \mathbf{k}}| \wedge d |{n, \mathbf{k}} \rangle
\end{equation}
The integral of the Berry curvature over the Brillouin torus  gives the first Chern number,
\begin{equation}
   c_1^n =   \int_{\mathbb{T}^2} f^n =  \int_{\mathbb{T}^2}  dk^2
   \left( \frac{\partial a^n_y}{\partial k_x} -  \frac{\partial a^n_x}{\partial k_y} \right)
 \end{equation}
For the Bloch bundle $\Bloch \rightarrow \mathbb{T}^2$, the total first Chern number
is then the sum
\begin{equation*}
  c_1 = \sum_{n=1}^N c_1^n \in \mathbb{Z}
\end{equation*}

The generalized Bloch--Hamiltonian for the quantum Hall effect (QHE) in a magnetic field in the Landau gauge given by $(0,eBx)$ is
\cite{TKNN82}:
\begin{equation}
\hat H(k_1,k_2)=\frac{1}{2m}(-i\hbar \frac{\partial}{\partial x}+\hbar k_1)^2+\frac{1}{2m}(-i\hbar \frac{\partial}{\partial y}+\hbar k_2-e Bx)^2+U(x,y)
\end{equation}
with the potential
$$
U(x,y)=U_1 \cos(2\pi x/a) + U_2\cos(2\pi y/b)
$$
and boundary conditions
$u_{k_1k_2}(x+qa,y)\exp(-2\pi i py/b)= u_{k_1k_2}(x,y+b)=u_{k_1,k_2}$
at rational flux $p/q$, see \cite{TKNN82} for details.
In particular the action of $\mathbb{Z}^2$ is restricted to a bigger
sub-lattice, such that the non--commutativity is lifted.
As was shown in \cite{TKNN82} and in the  following interpretations \cite{Simon,BES94} the quantization of the Hall conductance can be explained as the Chern class of a line bundle as discussed above. 
Fixing such a line bundle, for instance with the first Chern number $n=1$ is frequently called a quantum Hall state.

If this type of state is put next to an insulator, which has first Chern number $n=0$ then at the edge there are localized chiral edge modes \cite{HK10}.
This produces so called helical edge states as these states propagate towards the edge, where they get reflected, since there is a different topology in the neighboring bulks, but cannot penetrate deep into their own bulk, 
again because of the topological obstruction, see Figure 2 of \cite{HK10}.

\subsection{Dirac Hamiltonian}
\label{Diracsec}
The quantum Hall system can be viewed as a topological insulator, whose effective Hamiltonian is given by the Dirac Hamiltonian.  A phase transition is then controlled by some parameters in the Dirac Hamiltonian.
In order to illustrate such   phenomenon, we analyze  2--band models  generalizing the monopole (see below) that depend on a function $\mathbf{h}:B\to \mathbb{R}^3$, where $B$ is usually the Brillouin zone and in our case, 
to be concrete, the Brillouin torus $\mathbb{T}^2$. An element $\mathbf{k }= (k_x, \, k_y) \in \mathbb{T}^2 $ is called a crystal wave vector.
  A particular choice of $\mathbf{h}$, which we discuss, exhibits a phase transition and will also appear again in the study of the quantum spin Hall effect.

Consider the following Hamiltonian depending on $\mathbf{k}\in \mathbb{R}^3$:
\begin{equation}
H=\mathbf{k}\cdot \boldsymbol{\sigma}
 \end{equation}
 where the $\sigma_i$ are the Pauli matrices.
This Hamiltonian is non--degenerate outside $\mathbf{k}=0$.
Thus restricting it to $S^2\subset \mathbb{R}^2$ we are in the situation of the previous subsection, with the Bloch bundle being the sum of two line bundles $L_1\oplus L_2$. It is a nice calculation
to show that the Chern numbers are $\pm 1$ in this case \cite{Simon, B84}.
The degenerate point at $\mathbf{k}=0$ is thought of as a monopole
which is responsible for the ``charges'' given by the first Chern numbers.

There is a generalization to a $N$--band model, where now the $\sigma_i$ are spin $S=\frac{N-1}{2}$ matrices. In this case the  Chern numbers are  $n=2m$ with $m=-S,-S+1,\dots, S$ \cite{Simon, B84}.

More generally, consider the 2--level Hamiltonian
\begin{equation}
H
=h_x(\mathbf{k})\sigma_x+h_y(\mathbf{k})\sigma_y+h_z(\mathbf{k})\sigma_z
= \mathbf{h}(\mathbf{k}) \cdot \boldsymbol{\sigma}
\end{equation}
 Here $\boldsymbol{\sigma}=(\sigma_x,\sigma_y,\sigma_z)$, where the $\sigma_i$ are the Pauli matrices.

Assume that $H$ is nowhere degenerate, i.e.,\ $h_i(\mathbf{k})\neq 0$, the first Chern number in this case is a winding number. This goes back to \cite{B84} and is explained in detail in \cite{Lis14} Appendix A.
To see this, define the function:
\begin{equation*}
   \mathbf{d}: \mathbb{T}^2 \rightarrow S^2, \quad \mathbf{d}(\mathbf{k}) = \frac{\mathbf{h}(\mathbf{k})}{|\mathbf{h}(\mathbf{k})|}
\end{equation*}
The winding number of $\mathbf{d}$ computes the first Chern number of the lowest band, aka.\ the ground state.
\begin{equation}\label{monoc1}
   c_1 = \frac{1}{4\pi} \int_{\mathbb{T}^2} (\partial_{k_x}  \mathbf{d} \times \partial_{k_y}  \mathbf{d}) \cdot   \mathbf{d} \; d\mathbf{k}
\end{equation}
This is  the pull back via $\mathbf{d}$ of the 2-sphere ${S}^2$-- winding number around the monopole of $H=\mathbf{k}\cdot \boldsymbol{\sigma}$ at $\mathbf{k}=(0,0,0)$, see \cite{B84,S83}. This also works for  general Brillouin zones.
The first Chern number of the other band is then $-c_1$ since
the sum of the two bands is a trivial rank 2 bundle.

The particular choice of $\mathbf{h}(k_x,k_y)$ yields:
\begin{equation}
   H(\mathbf{k}) = A \sin (k_x) \sigma_x + A \sin (k_y) \sigma_y + [M - B ( \sin^2({k_x}/{ 2})  + \sin^2 ({k_y}/{2}) ) ] \sigma_z
\end{equation}
where
 $A, B, M$ are parameters.
$M$ can be viewed as a mass parameter, and $B$ plays the role of a magnetic field.
This Hamiltonian appears in the analysis of an electron
on a 2d square lattice with spin-orbit interaction.

It has the first Chern number of the lowest band determined
by the signs of parameters \cite{SSL11},
\begin{equation}
   c_1 = -\frac{1}{2} [sgn(M) + sgn(B)]
\end{equation}

Using the usual first order approximation as $\mathbf{k} \rightarrow 0$, and setting $B'=\frac{1}{4}B$,  the Hamiltonian takes the form:

\begin{equation}\label{dirhameq}
   H (\mathbf{k})= A k_x \sigma_x + A k_y \sigma_y + [M - B'(k_x^2 + k_y^2)]\sigma_z
\end{equation} which is commonly called a Dirac Hamiltonian.
Note that comparing this expression to a Dirac operator, the above Hamiltonian has a quadratic correction term $B' \mathbf{k}^2 \sigma_z$.
It also appears in the analysis of the quantum spin Hall effect in the quantum well, see \S\ref{QSHpar}.

This system depends heavily on the relation between $M$ and $B$, since this determines the coupling to $\sigma_z$. I.e.\ $M$ and $B$  determine the locus of $ M - B\mathbf{k}^2 = 0$.
If we change from $B > M$ to $B < M$, then the locus will change accordingly, which gives rise to a phase transition \cite{SSL11}.
Using this point of view, the quantum Hall system manifests itself as  the simplest topological insulator, as we will now discuss.

\subsection{Bulk-edge correspondence} \label{BEcorr}

We have seen that the first Chern number characterizes a bulk state in the quantum Hall effect. When considering
  an interface  with an insulator, an edge state appears. In order to better understand this phenomenon, we consider two copies of the model (\ref{dirhameq}), but with different ground state Chern numbers, 
  put next to each other along an interface.
This is an example of a topological insulator with a domain wall.
We discuss how the connection between edge and bulk states
can be understood in terms of an index theorem.

 For this we simplify the situation to consider a two--band system. This is enough to capture the phenomenon as we only need to consider two bands separated by a gap in which the Fermi level lies. 
 The Hamiltonians are then of Dirac Hamiltonian type and are generalized to include terms that may depend on higher derivatives of $x$ and $y$, see below for an example and \cite{FSFF12} for details.

The generalized index theorem for this type of Hamiltonian connects the analytical index (aka.\ the spectral flow) of the edge states and the topological index (aka.\ the first Chern number) of the bulk.
The spectral flow of the edge states and the variation of first Chern numbers specifying the bulks establishes a so--called bulk-edge correspondence.

We illustrate this for  the Dirac Hamiltonian (\ref{dirhameq}), whose position representation is:
\begin{equation}
   H_b = -i\partial_x \sigma_x -i \partial_y \sigma_y + (M + B\partial_x^2 + B \partial_y^2)\sigma_z
\end{equation}
where the subscript $b$ stands for  ``bulk Hamiltonian''. As mentioned previously, the corresponding edge Hamiltonian is obtained by a partial
Fourier transformation in the $y$--direction; see end of \S \ref{landauedgepar}.
\begin{equation}
   H_e = -i\partial_x \sigma_x +k_y \sigma_y + (M + B\partial_x^2 - B k_y^2)\sigma_z
\end{equation}

One  introduces a domain wall  at $x = 0$  separating two bulk systems, by letting
the mass term become dependent on $x$.  This  $M(x)$ is assumed to have stabilized limits,
 $\lim_{x \rightarrow \pm \infty}M(x) = M_{\pm}$.
The edge eigenstates are assumed to satisfy the eigenvalue equation of the edge Hamiltonian
\begin{equation*}
   H_e(x,k_y)\phi_n(x,k_y) = \varepsilon_n(k_y) \phi_n(x,k_y)
\end{equation*}

We make the usual assumption, that as $|k_y|$ gets big, then eventually $\varepsilon_n(k_y)$ will lie in the conduction or valence band.
An  edge eigenstate $\phi_n$ is called an edge state, if it connects the valence band and the conduction band. This means that as one varies $k_y$, $\varepsilon_n(k_y)$ hits both the valence and the conduction band.
Hence for any edge state  $\phi_n(k_y)$, $\varepsilon_n(k_y)$ intersects with the Fermi level $E_F =0$ (after a possible shift).

We let $sf(\phi_n)$ be the intersection number of $\varepsilon_n(k_y)$ with $E_F=0$, i.e.\ the spectral flow of the edge state $\phi_n$.
If the intersection is transversal and there are only finitely many points of intersection, i.e.\ points where $\varepsilon_n(k_y)= 0$, then the spectral flow of the edge state $\phi_n$ is
\begin{equation}
   sf(\phi_n) = n_+ - n_- \in \{-1,1\}
\end{equation}
where $n_+$ is the number of times that the sign of $\varepsilon_n$ changes from negative to positive and
$n_-$ is the number of times that the sign of $\varepsilon_n$ changes from positive to  negative when increasing $k_y$. Note that $sf$ makes sense for all the edge eigenstates, but is $0$ for the ones that are not edge states.

Set
$$
\nu=\sum_n sf(\phi_n)=N_+-N_-
$$
where the sum is either over all eigenstates or over the edge states,
$N_+$ is the number of edge states going from the valence to the conduction band as $k_y$ increases and $N_-$ being the number of states going from the conduction to the valence band.
In order to identify the spectral flow with an analytical index, one introduces an
extended Dirac Hamiltonian \cite{FSFF12}
\begin{equation}
   \tilde{H}_e = -iv \tau_1\partial_{k_y} + \tau_2 H_e
\end{equation}
where $\tau_i$ are the Pauli matrices and $v$ is a constant parameter representing a velocity. This is a block $4\times 4$ matrix Hamiltonian with the diagonal blocks being zero. 
The anti--diagonal blocks represent the two processes corresponding to $N_+$ and $N_-$. In this form it looks like the Dirac operator
in the Atiyah-Patodi-Singer index theorem \cite{APS76}.
Based on the chiral symmetry of $\tilde{H}_e$
($\{\tilde H_e,\tau_3\}=0$) one can compute that
the analytical index of $\tilde{H}_e$ equals the negative of the spectral flow of the edge states,
\begin{equation*}
   ind_a( \tilde{H}_e) = - \nu
\end{equation*}

The topological index  of $\tilde{H}_e$ was also computed explicitly \cite{FSFF12} by investigating  the divergence of a chiral current of $\tilde{H}_e$, see {\it loc.\ cit.} for details.
The topological index turns out to be the difference of the  first Chern numbers specifying the two bulk ground states
separated by the domain wall,
\begin{equation*}
   - ind_t(  \tilde{H}_e) = c_1(M_+ ) - c_1(M_-)
\end{equation*}

By the Atiyah--Patodi--Singer index theorem, the analytical index equals the topological index for $\tilde{H}_e$ and hence
 by  the generalized index theorem
\begin{equation}
   \nu = c_1(M_+ ) - c_1(M_-)
\end{equation}
which can be interpreted as the bulk-edge correspondence  between the two bulk states and the edge states. A
physical picture is as follows:  by stacking two systems with topologically distinct ground states, an edge state is created at the interface.  
The transport property of the edge state is then determined by the topology of the junction. This would also work for  edge states in the quantum spin Hall effect, see e.g.\ \cite{HK10}.

\section{$\mathbb{Z}_2$ topological insulators} \label{Z2Inv}
In this section,  we review  the different definitions of the $\mathbb{Z}_2$ invariant for time reversal invariant free (or weak-interacting) fermionic topological insulators.

For 2d and 3d  time reversal invariant topological quantum systems, $\Z_2$ invariants characterizing topological insulators were introduced in \cite{KM05,FK06,FKM07,MB07}.
We will first review their definitions in this section. There are actually several natural definitions for these, using Pfaffians, polarizations,  determinant line bundle and holonomy etc. 
Besides giving these definitions, we explain how they are equivalent.

Physically, the $\Z_2$ invariant can be interpreted as the parity  of so--called Majorana zero modes.
In particular calling the invariant $\nu$, for a 2d quantum spin Hall system, $\nu \equiv 0$ (resp. $\nu \equiv 1$) when there exist even (resp. odd) pairs of helical edge states.
Note that one helical edge state consists of two chiral edge states.
Similarly for 3d time reversal invariant topological insulators, $\nu \equiv 0$ (resp. $\nu \equiv 1$)
when there exist even (resp. odd ) numbers of Dirac cones, i.e.,
conical singularities produced by surface states  at the fixed points of time reversal symmetry.

We will go through the details in several subsections.
 \S \ref{background} is an introduction to time reversal symmetry and $\mathbb{Z}_2$ topological insulators. We provide examples of two 2d and one 3d topological insulators.
 We then go on to define the 2d Kane--Mele invariant in its original version using Pfaffians and determinants, being careful to define the setup rigorously. We also discuss its 3d generalization in \S \ref{KMinvt}.
Here we first consider the general structure of Bloch bundles with time reversal $\mathbb{Z}_2$ symmetry.
The next subsection \S \ref{Quatbundle} then gives a geometric construction of the $\mathbb{Z}_2$ invariant using bundle theoretic means. While \S \ref{homotopypar} 
reviews the homotopy theoretic characterization of the $\Z_2$ invariant that leads over to the considerations of \S \ref{indexpar}.

\subsection{Background and examples}\label{background}
We first recall the time reversal symmetry, which introduces  Kramers degeneracy and changes the effective topology and geometry of
the game.
Next, we review the effective  Hamiltonians of $\mathbb{Z}_2$ topological insulators, which can be used to  define elements in K-homology in the future.

\subsubsection{Time reversal symmetry}

A physical system has time reversal symmetry  if it is invariant under the time reversal transformation $\mathcal{T}: t \mapsto -t$.  In quantum systems the
time reversal symmetry is represented by a time reversal operator $\Theta$. By general theory \cite{Wigner}, $\Theta$ is necessarily anti--unitary and can be realized as the product of a unitary operator and the
complex conjugation operator. Again by general theory, in momentum space (or the Brillouin zone)
$\mathcal{T}$ acts as $\mathbf{k} \mapsto -\mathbf{k}$.

A time reversal invariant model is required to have $[H(\mathbf{r}), \Theta] = 0$, or in the momentum representation
\begin{equation}
\label{Hequieq}
   \Theta H(\mathbf{k}) \Theta^{-1} = H(- \mathbf{k})
\end{equation}

For example, in a two-band system the time reversal operator $\Theta$ is defined by
\begin{equation}
   \Theta = i  \sigma_y K
\end{equation}
where $\sigma_y$ is the imaginary Pauli matrix and $K$ is the complex conjugation. $\Theta$ is indeed an anti-unitary operator, i.e., for physical states $\psi$ and $\phi$,
\begin{equation}
\label{orthoeq}
  \langle \Theta \psi, \Theta \phi \rangle =  \langle  \phi, \psi \rangle, \quad \quad \Theta (a \psi + b \phi) = \bar{a} \Theta\psi + \bar{b} \Theta \phi, \quad a, b\in \mathbb{C}
\end{equation}
For a spin-$\frac{1}{2}$ particle, such as an electron, it also has the property:
\begin{equation}
   \Theta^2 =-1
\end{equation}
This results in the so--called Kramers degeneracy, which is the fact that  all energy levels are doubly degenerate in a time reversal invariant electronic system with an odd number of electrons. 
In fact, $\phi$ and $\Theta(\phi)$ are orthogonal:

\begin{equation}
\label{Wignereq}
\langle \Theta\phi,\phi\rangle= \langle \Theta\phi, \Theta\Theta\phi\rangle=-\langle\Theta\phi,\phi\rangle=0
\end{equation}
In general, $\Theta$ is skew--symmetric in the sense that
\begin{equation}
\label{skeweq}
\langle \Theta\psi,\phi\rangle= \langle \Theta\phi, \Theta\Theta\psi\rangle=-\langle\Theta\phi,\psi\rangle
\end{equation}

The Dirac Hamiltonian
$ H(\mathbf{k}) =  k_x \, \sigma_x +  k_y \, \sigma_y + (M - B\mathbf{k}^2) \sigma_z$
is not time reversal invariant.
In contrast, the models for quantum spin Hall effect  with spin-orbit coupling given in the next section are time reversal invariant systems.

\subsubsection{Effective Brillouin zone}
\label{EBZpar}
\begin{figure}[t]
 \begin{center}
 \includegraphics[height=3in]{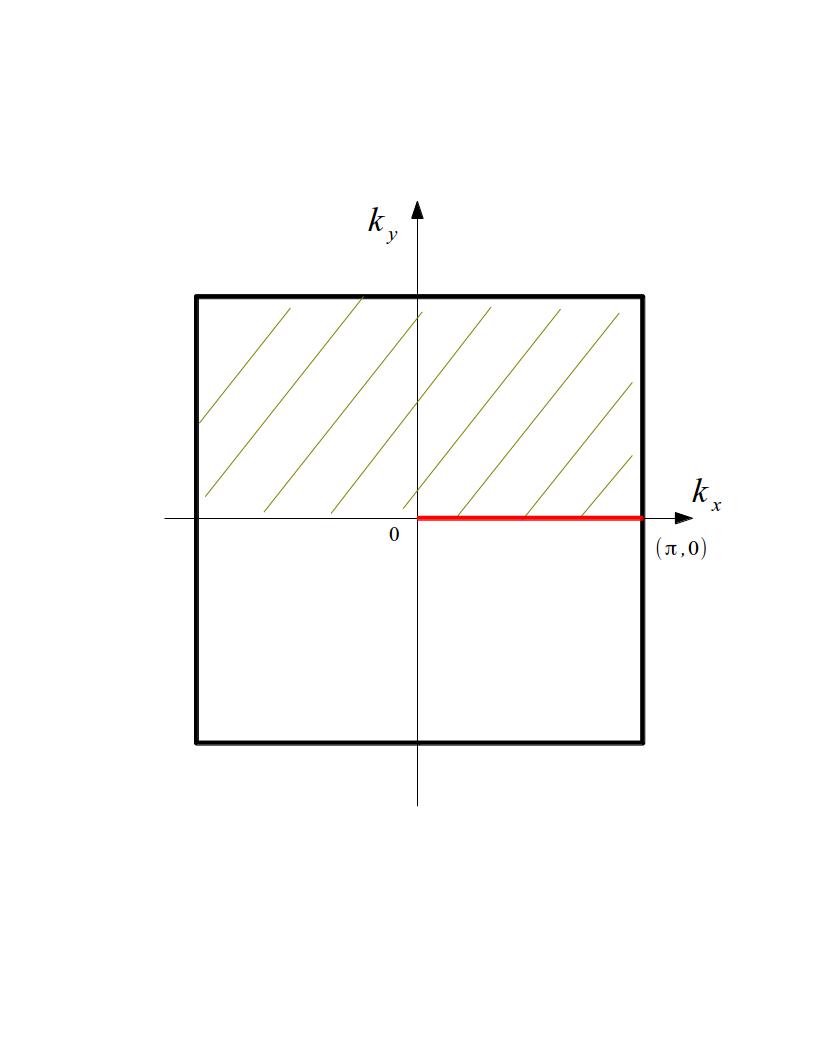}
 \includegraphics[height=3in]{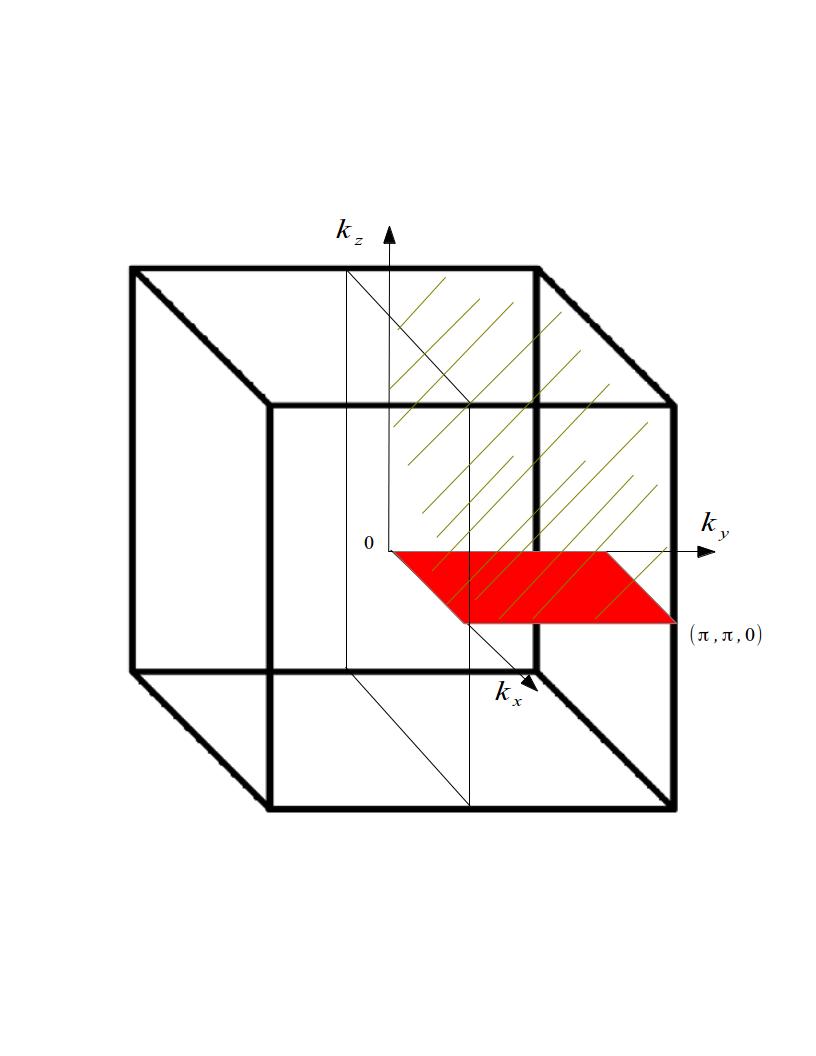}
 \end{center}
 \caption{Effective Brillouin zone in 2d and 3d}
\label{figEBZ}
\end{figure}

The association $\bk\to H(\bk)$ is formally a map $H:\TT^d\to Herm_n$, where $Herm_n$ is the space of Hermitian $n\times n$ matrices.
If we take into account the action of $\T$ on $\TT^d$, then to specify a map with time reversal symmetry it is enough to know the restriction of $H$ onto a fundamental domain of the $\T$ action, 
by (\ref{Hequieq}). Such a choice of fundamental domain will be called Effective Brillouin Zone (EBZ).
In the 2d--case such an effective Brillouin zone is given by a cylinder $C=S^1\times I\subset \TT^2$ and in 3d by $\TT^2\times  I \subset \TT^3$, see Figure \ref{figEBZ}.

If one considers the boundary of this EBZ, then it still has a non--trivial action of $\T$ and
we can consider a fundamental domain of the boundary of the EBZ, which we call $R^{d-1}$.
In physics parlance, this region characterizes
the edge states using a unique coordinate. For example, $R^1 = [0, \pi]$ in the 2d effective Brillouin zone $[0, \pi] \times [-\pi, \pi]$
and  $R^2$ is the square $ [0, \pi]\times [0, \pi]$ in the 3d case, see Figure \ref{figEBZ}.

Although the usual Brillouin zone is $\TT^d$, the spheres $S^d$ play a special role. First they can be viewed as local invariants as in Simon's interpretation of the TKNN integers \cite{S83} 
or generally as enclosing monopoles or higher codimension degeneracy loci.
For instance, from  \cite{vNW}, we know that for a family of Dirac Hamiltonians $H \in \mathfrak{H}$, its eigenvalue degeneracy is generically of
codimension three, so in 3d  for a generic family each  point $\bk$ corresponding to a degenerate eigenvalue, aka.\ monopole, is enclosed by an $S^2$.
Secondly, the spheres $S^d$ occur through effective maps as
in the Moore-Balents description below \S\ref{homotopypar}, or the more complicated arguments presented in \S\ref{genhompar}. Finally, they appear in the stable splitting, see \S\ref{Ktheorypar}. 
In particular, they arise through pull-back via the collapse map $q:\TT^d\to S^d$, see {\it loc.\ cit.}.

Thus, we will look at maps $H:X\to Herm_n$ where $X$ can be $\TT^d,S^d, $ or $R^{d-1}$. Depending on the situation, the target may be a subspace of all Hermitian matrices, see \S \ref{homotopypar}.

\subsubsection{Quantum spin Hall effect}
\label{QSHpar}
We now turn to the  Hamiltonians of the quantum spin Hall effect, which appears in both graphene and the
HgTe quantum well. These are  2d time reversal invariant models.
Haldane \cite{H88} first investigated the quantum Hall effect in graphene which has a honeycomb lattice structure.
The Kane--Mele model \cite{KM0501} further took spin-orbit coupling into account and generalized the spinless Haldane
 model. They first predicted the quantum spin Hall effect in graphene with the effective Dirac Hamiltonian
\begin{equation}
   H_{KM} = \sum_{i=1}^5 {d}_i(\mathbf{k}) {\Gamma}_i + \sum_{1=i < j}^5 {d}_{ij}(\mathbf{k}) {\Gamma}_{ij}
\end{equation}
where for $i=1,\dots,5$ the   gamma matrices $\Gamma_i$ are defined via
\begin{equation*}
   \boldsymbol{\Gamma} = (\sigma_x \otimes s_0, \sigma_z \otimes s_0, \sigma_y \otimes s_x, \sigma_y \otimes s_y, \sigma_y \otimes s_z )
\end{equation*}
where $\sigma_i$ are as before and the $s_i$ are Pauli matrices as well, representing the electron spin. Their commutators define the $\Gamma_{ij}$:
\begin{equation*}
   \Gamma_{ij} = \frac{1}{2i} [\Gamma_i, \Gamma_j]
\end{equation*}

The time reversal operator in this model is  defined by
\begin{equation}
\label{4bandtheta}
    \Theta = i (\sigma_0 \otimes s_y) K
\end{equation}
It is straightforward to check that
\begin{equation*}
   \Theta \Gamma_i \Theta^{-1} = \Gamma_i, \quad \Theta \Gamma_{ij} \Theta^{-1} = - \Gamma_{ij}
\end{equation*}
Combining this  with the properties of the coefficients \cite{KM0501},
\begin{equation*}
   d_i(- \mathbf{k}) = d_i(\mathbf{k}), \quad d_{ij} (-\mathbf{k}) = - d_{ij}(\mathbf{k})
\end{equation*}
one obtains that the Hamiltonian $H_{KM}$ is time reversal invariant.

The coefficients also depend on a parameter, which is called the nearest neighbor Rashba coupling constant.
If the Rashba coupling constant vanishes, then the  Hamiltonian above is invariant under spin rotations along the z-axis,
i.e., $[H_{KM}, s_z] = 0$. Then it splits into two components for the up and down spins,
\begin{equation*}
   H = \begin{pmatrix}
        H_{\uparrow} (\mathbf{k}) & 0\\
        0& H_{\downarrow}(\mathbf{k})
       \end{pmatrix}
\end{equation*}
where
\begin{equation}
   \begin{array}{ll}
     H_{\uparrow} = d_1 \sigma_x - d_{12}\sigma_y + (d_2 - d_{15}) \sigma_z \\
        H_{\downarrow}= d_1 \sigma_x - d_{12}\sigma_y + (d_2 + d_{15}) \sigma_z
   \end{array}
\end{equation}
This system is basically comprised of two copies of the Haldane model.
Each component from the above defines a quantum Hall system, which is characterized by its first Chern number.
The total Chern number of the Kane--Mele model is always zero, since  for a time reversal invariant ground state $ c_{1\uparrow} =- c_{1\downarrow} $ and hence
\begin{equation*}
   c_1 = c_{1\uparrow} + c_{1\downarrow} =0
\end{equation*}
It is, however,  possible to define a new topological invariant, i.e., the spin Chern number
\begin{equation}
   2c_{spin} :=  c_{1\uparrow} - c_{1\downarrow}
\end{equation}

In general, the Rashba coupling is non-zero, then the spin-up and spin-down components are coupled together and the definition of the
 spin Chern number is no longer valid. However,
there always exists
a $\mathbb{Z}_2$ invariant for the quantum spin Hall effect, for example, the Kane-Mele invariant discussed in section \S\ref{Z2invsec}.

If a topological insulator can be
specified by a $\mathbb{Z}_2$ invariant, we call it a $\mathbb{Z}_2$ topological insulator.

The quantum spin Hall effect was also predicted by Bernevig, Hughes and Zhang \cite{BHZ06} for the
HgTe/CdTe quantum well with stronger spin-orbit interaction, which was  observed by K\"{o}nig et al. \cite{K07} soon after the prediction.
The effective Hamiltonian for this so--called quantum
well is given by
\begin{equation*}
   H_{BHZ} = \begin{pmatrix}
        H(\mathbf{k}) & 0\\
        0& H^*(-\mathbf{k})
       \end{pmatrix}
\end{equation*}
where
\begin{equation}
    H(\mathbf{k}) =  Ak_x \sigma_x + Ak_y \sigma_y + (M- B\mathbf{k}^2)\sigma_z + (C-D\mathbf{k}^2)\sigma_0
\end{equation}
here $A,B,C,D$ are expansion parameters and $M$ is a mass or gap parameter.  Notice that this effective Hamiltonian, up to the $\sigma_0$--term ---which actually drops out in the quantum Hall response---, 
is of the form discussed in \S\ref{Diracsec}.
This system again decouples, so that the simpler version of
the $\Z_2$ invariant given by $c_{spin}$ is applicable.

\subsubsection{3d topological insulators}

Later on, 3d topological insulators were also observed in nature.   In 3d the role of edge states is played by codimension--1 surface states. 
The observed surface states look like double cones in the spectrum or energy in the momentum dispersion relation in physical terms. 
They  take the form of local conical singularities $x^2 + y^2 = z^2$, and are called Dirac cones by physicists, see e.g.\ \cite{KKWKsing}. 
A point in the Brillouin torus over which  a Dirac cone lies is called a Dirac point.
In 3d topological insulators, a Dirac cone is sometimes also referred to as a Majorana zero mode.
Note that a Majorana zero mode should not be confused with a Majorana fermion.

 If the system is invariant under time reversal symmetry then the so--called Kramers degeneracy forces level sticking at  any fixed point of the time reversal symmetry,  as discussed in the next section.
 If these singularities are not extended and are conical one has  Dirac cones.  The number of Dirac cones determines whether the system is a topological insulator, that is, if a material has odd number of Dirac cones,
then it is a stable 3d topological insulator.

For example, the effective Dirac Hamiltonian for $Bi_2Se_3$ with spin orbit coupling has a single Dirac cone \cite{ZZ09}. The Hamiltonian is
\begin{equation*}
   H(\mathbf{k}) =    A_1k_x \sigma_x \otimes s_x
                    + A_1k_y \sigma_x \otimes s_y + A_2k_z \sigma_x \otimes s_z + M(\mathbf{k}) \sigma_z \otimes s_0 +\epsilon_0(\mathbf{k}) I_4
\end{equation*}
where $I_4$ is the $4\times 4$ identity matrix,
\begin{equation}
  \begin{array}{ll}
   M(\mathbf{k})= M - B_1 k_z^2 - B_2 (k_x^2 + k_y^2) \\
     \epsilon_0(\mathbf{k}) = C + D_1 k_z^2 + D_2 (k_x^2 + k_y^2)
  \end{array}
\end{equation}
and  the $A_i,B_i,C, D_i$ and $M$ are model parameters.

\subsection{Kane--Mele invariant}\label{KMinvt}
In this section, we carefully present the relevant structures that the time reversal symmetry induces on a Bloch bundle and then use these structures to define the Kane--Mele invariant in 2d and 3d, 
following their original construction  and
finish with a holonomy interpretation of the invariant.
 We will also discuss a coordinate free bundle theoretic description of these constructions based on canonical orientations in the following section.

\subsubsection{Kramers degeneracy}
In this subsection, we  discuss  the geometric setup for time reversal invariant topological insulators.
We are following  physical notations, this means that we use sections. These are allowed to be non--zero only locally, or in other words locally defined. So we have to stress that in order to get an global object,
sometimes extra work is needed to glue local sections together.

Consider a time reversal invariant Dirac Hamiltonian $H(\mathbf{k})$, $\mathbf{k} \in \mathbb{T}^d$,
whose eigenvalue equation is $H(\mathbf{k}) u_n(\mathbf{k}) = E_n(\mathbf{k}) u_n(\mathbf{k}) $.
 For each physical state $| u_n(\mathbf{k}) \rangle = |n, \mathbf{k} \rangle$, $\Theta | n, \mathbf{k} \rangle$ gives another state with the same energy,
and such doublet  is called a Kramers pair.
At a time reversal invariant point $\mathbf{k}= \mathcal{T}(\mathbf{k})=-\mathbf{k}\in \mathbb{T}^d$ time reversal symmetry forces degeneracy or a level crossing. That is
both $|u_n(\mathbf{k})\rangle$ and  $\Theta|u_n(\mathbf{k})\rangle$
are states with momentum $\mathbf{k}=\mathcal{T}(\mathbf{k})=\mathbf{-k}$ and the same energy. They are linearly independent, due to (\ref{Wignereq}). This is called Kramers degeneracy.

Starting with a lattice model, in the reciprocal lattice the time reversal invariant points are given by
\begin{equation*}
   -\mathbf{\Gamma}_i = \mathbf{\Gamma}_i + \mathbf{n}_i \cdot \mathbf{K}, \quad \text{i.e.} \quad \mathbf{\Gamma}_i = \frac{1}{2}\mathbf{n}_i \cdot \mathbf{K},
\end{equation*}
where $\mathbf{K}$ is the reciprocal lattice vector and the coordinates of $\mathbf{n_i}$ are $0$ or $1$. For example, in a 2d square lattice
with $\mathbf{K} = (2\pi, 2\pi) $, we have four time reversal invariant points $\mathbf{\Gamma} = \{ (0, 0), (0, \pi), (\pi, 0), (\pi, \pi)\}$.
So, for instance, the Brillouin torus $\mathbb{T}^2$ has four fixed points and similarly $\mathbb{T}^3$ has eight fixed points of the time reversal symmetry.

Hence in such a system, the Bloch bundle does not split into line bundles according to the energy eigenvalues, but under the customary assumption that the Kramers degeneracies are the only degeneracies,
the Bloch bundle does split into a direct sum of rank 2 vector bundles, as we now discuss.
We  assume from now on  that  indeed {\em all} degeneracies that are present are forced by Kramers degeneracy.
Physically, this can for instance be due to a spin--orbit coupling.
First, let's assume that the Bloch bundle has even rank $2N$, so that there are $2N$ states, which at the fixed points form pairs. Near a fixed point, the pairs may, and under the assumption will, split.
  So, in a neighborhood of  a fixed point, each Kramers pair will give rise to two energy levels, spanned by say,  $|u_{2n-1}  (\mathbf{k})\rangle, |u_{2n}( \mathbf{k})\rangle$. We will use the notation:
\begin{equation}
\label{paireq}
   |u_n^1(\mathbf{k})\rangle= |u_{2n-1}( \mathbf{k}) \rangle, \quad
    |u_n^2(\mathbf{k})\rangle =|u_{2n}( \mathbf{k}) \rangle
\end{equation}

 In this case, if there are only finitely many occupied states, there will be no crossings between levels from different pairs (\ref{paireq}).\footnote{A quick proof goes as follows. 
 Suppose that for some $n$ the levels  $2n$ and $2n+1$ cross. Then this must happen at a time reversal invariant point. This implies that at this point the levels $2n$ and $2n+1$ cross for all $n$. 
 This will leave the first levels  and last levels unpaired. Notice that in an infinite system, say indexed by $\Z$, such a cross over from the pairing $2n-1,2n$ to $2n,2n+1$ can happen. 
 This observation was the initial  spark for the definition of the Kane--Mele $\Z_2$ invariant and is pertinent to the  discussion of the edge state and the bulk--boundary correspondence.} 
 This implies that if the Bloch bundle has even rank $2N$ it can be split
using the Kramers doublets to  define  rank two sub--bundles
$$\V_n \rightarrow \mathbb{T}^d \quad \text{s.t.} \quad \Bloch = \oplus_{n=1}^N \V_n$$

By the same type of argument if the Bloch bundle has rank 2N+1, then it splits into rank two bundles and one line bundle. We will not consider this case further as one can split off the line bundle from the Bloch bundle.

The  bundle
 $\Bloch$ and the sub-bundles $\V_n$ also carry the operation of $\Theta$ which is compatible with the operation $\mathcal{T}$ on the base. As we will discuss in \S \ref{bundletypepar} 
 this extra structure makes $\Bloch$ and the $\V_n$ into  so--called Quaternionic vector bundles.
Sometimes, we will wish to emphasize this and then we will write $\BlochT = \oplus_{n=1}^N \BlochT_n$. Notice that $\Theta$ induces an isomorphism $\mathcal{B}\simeq \mathcal{\bar \Bloch}$.
More precisely we have that $\mathcal{T}^*\Theta(\mathcal{V}_n)\simeq\bar{\mathcal{V}}_n$.

One consequence of this is that all the odd Chern classes of
$\BlochT$ are two--torsion
$2c_{2i+1}(\Bloch)=0\in H^{4i+2}(\mathbb{T}^d,\Z)$,
since $\Bloch\simeq \bar{\Bloch}$ implies that $c_{i}(\Bloch)=c_{i}(\bar{\Bloch})=(-1)^{i}c_{i}(\Bloch)$.   Hence
they vanish when extending the coefficients to $\R$ as is done under the deRham isomorphism or if the cohomology is torsion free. Real coefficients are always used in the computation of Chern classes via integrals, i.e.\ Chern--Weil theory.
 In particular this shows that the $c_1(\mathcal{B})=0\in H^2({\mathbb T}^d,\R)$ for instance when computed using integrals over curvature or holonomy.
 Thus the first non--trivial information will be contained in the second Chern class $c_2(\Bloch)\in H^4(\TT^d)$ or more generally the $H^4$ of the base of the bundle. 
 Indeed,  for 4-dimensional time reversal invariant models the fundamental topological invariant is the second Chern number \cite{QHZ08}.
However, $H^4$ is possibly non--zero only for manifolds of dimension greater or equal to $4$, so it is not useful directly in 2d and 3d.
 This is the reason for the introduction of the $\Z_2$ invariant. For 3d one can use dimensional reduction from 4d to 3d to  understand
 how the $\Z_2$ invariant appears as the dimensional reduction of $c_2$.

The vanishing of $c_1(\mathcal{B})$, however, also implies that the first Chern class of its determinant line bundle vanishes, i.e., $c_1(Det \Bloch)=0\in H^2({\mathbb T}^d)$, since the cohomology of $\mathbb{T}^d$ is torsion free. 
In addition, this determinant line bundle has a global section, which is one of the main ingredients of the $\mathbb{Z}_2$ invariant.
The main key to understanding the $\Z_2$ invariant  geometrically can be traced  back to  \cite{FK06}, which we now make precise.

\subsubsection{Geometric setup}
\label{KMprep}

One rough idea about how to obtain the invariant is given as follows.
At any fixed point $\mathbf{k}_0$, $\mathbf{k}_0=-\mathbf{k}_0$,
$\Theta$ is a skew--symmetric (\ref{skeweq}) anti--linear morphism on the fiber over $\bk$. {\it After choosing a basis}, a natural object to study is the Pfaffian of the resulting skew--symmetric matrix. 
Since we can reduce to rank 2 bundles this is just given by the top right corner of a $2\times2$ matrix.
The actual choice of basis plays a role, as the sign of Pfaffian changes if for instance the two basis vectors are permuted. 
The Kane-Mele invariant is a way to normalize and compare these choices of Pfaffians at all fixed points simultaneously. For this one
needs to make a coherent choice of normalization and show that the result is independent of the choice, which is what the invariant does. 
A key  fact that is used is that the determinant line bundle has global sections. This can be seen in several different ways as we now discuss.

To study the choice of basis, first observe that at any fixed point $\mathbf{k}_0$, $\mathbf{k}_0=-\mathbf{k}_0$, $\Theta |\phi\rangle$ and $ |\phi\rangle$ are linearly independent by (\ref{Wignereq}), so that we can choose
$\Theta  |u_n^1(\mathbf{k}_0)\rangle$ and $\Theta  |u_n^2(\mathbf{k}_0)\rangle$, s.t.\
$$\Theta  |u_n^1(\mathbf{k}_0)\rangle= |u_n^2(\mathbf{k}_0)\rangle
\text{ and } \Theta  |u_n^2(\mathbf{k}_0)\rangle= -|u_n^1(\mathbf{k}_0)\rangle$$
where the second equation follows from the first by $\Theta^2=-1$.

In contrast if $|u_n^1(\mathbf{k})\rangle$ and $|u_n^2(\mathbf{k})\rangle$ are states, i.e.\ local sections, defined on a subset $V$ that is stable under $\T$, i.e.\ $\T(V)=V$, and contains {\bf no} fixed points, 
i.e.\ $V\cap Fix(\T)=\emptyset$, then, since $\Bloch$ is trivial over ${\mathbb T}^d \setminus Fix(\T)$ by the assumption that there are only degeneracies at the fixed points, and
 $\Theta$ commutes with $H$, we get that away from a fixed point
\begin{equation}
\label{chidefeq}
    u_n^1(-\mathbf{k}) = e^{i\chi^1_n(-\mathbf{k})} \Theta u_n^1(\mathbf{k})
\quad
       u_n^2(-\mathbf{k})  = e^{i\chi^2_n(-\mathbf{k})} \Theta  u_n^2(\mathbf{k})
\end{equation}
the relation $\Theta^2=-1$ implies that
$e^{i\chi^{1,2}_n(-\mathbf{k})}=-e^{i\chi^{1,2}_n(\mathbf{k})}$.

Together these two facts imply that  the $u_n^1$ do not extend
as continuous sections to the fixed points in any $\T$ stable $V$ that does contain fixed points. However, one can cobble together a section using both $u^1_n$ and $u_n^2$ on certain 1--dimensional domains,
for instance in the case of a 1--dimensional Brillouin zone.

In particular, let $V$ be a small $\Theta$ stable path through a fixed point $\mathbf{k}_0$. Decompose $V$ as $V^+\amalg V^-$ with $\T(V^+)=V^-$ and
$V^+\cap V^-=\mathbf{k}_0$.
Now we can fix sections $|u_n^1(\mathbf{k})\rangle$ and $|u_n^2(\mathbf{k})\rangle$ independently on $V^+$ and $V^-$, by continuing the states continuously to the fixed point $\mathbf{k}_0$ from both sides. 
By perturbation theory, we can choose these to match up if necessary after multiplying them with a local phase factor.
$$
\lim_{\k\to\k_0 }  u_n^1(\mathbf{k})=\lim_{-\k\to\k_0 }  u_n^2(\mathbf{-k})
\quad \text {and} \quad
\lim_{\k\to\k_0 }  u_n^2(\mathbf{k})=\lim_{-\k\to\k_0 }  u_n^1(\mathbf{-k})
$$

The reason for this is that after subtracting a constant term, the family is given by $H(\k)={\bf x}(\k)\cdot \boldsymbol{\sigma}$, with ${\bf x}(\k_0)=0$ and
${\bf x}(\k)$ an odd function of $\k$. This means that up to rotation, linearly near $\k_0$:
$H(\k)=(\k-\k_0)\sigma_z+O(\k^2)$, whence the claim follows by standard perturbation theory.
Here we need that we assumed that the fixed points are isolated.

These equations mean that one can glue two sections together over $V$, which we call $|u_n^I(\mathbf{k})\rangle$ and $|u_n^{II}(\mathbf{k})\rangle$ with
$|u_n^I(\mathbf{k})\rangle|_{V^+}=|u_n^1(\mathbf{k})\rangle, |u_n^I(\mathbf{k})\rangle|_{V^-}=|u_n^2(\mathbf{k})\rangle$ and
$|u_n^{II}(\mathbf{k})\rangle|_{V^+}=|u_n^2(\mathbf{k})\rangle, |u_n^{II}(\mathbf{k})\rangle|_{V^-}=|u_n^1(\mathbf{k})\rangle$. Likewise, in this case, 
we can glue together the functions $\chi^1_n(-\k)$ and $\chi^2_n(\k)$ at $\k_0$ to form one function $\chi_n(k)$. For these sections, we obtain that

\begin{equation} \label{rel1}
    u_n^{II}(-\mathbf{k}) = e^{i\chi_n(-\mathbf{k})} \Theta u_n^I(\mathbf{k})
\text{ and }       u_n^I(-\mathbf{k})  = -e^{i\chi_n(\mathbf{k})} \Theta  u_n^{II}(\mathbf{k})
\end{equation}

\begin{figure}[t]
 \begin{center}
 \includegraphics[height=2in]{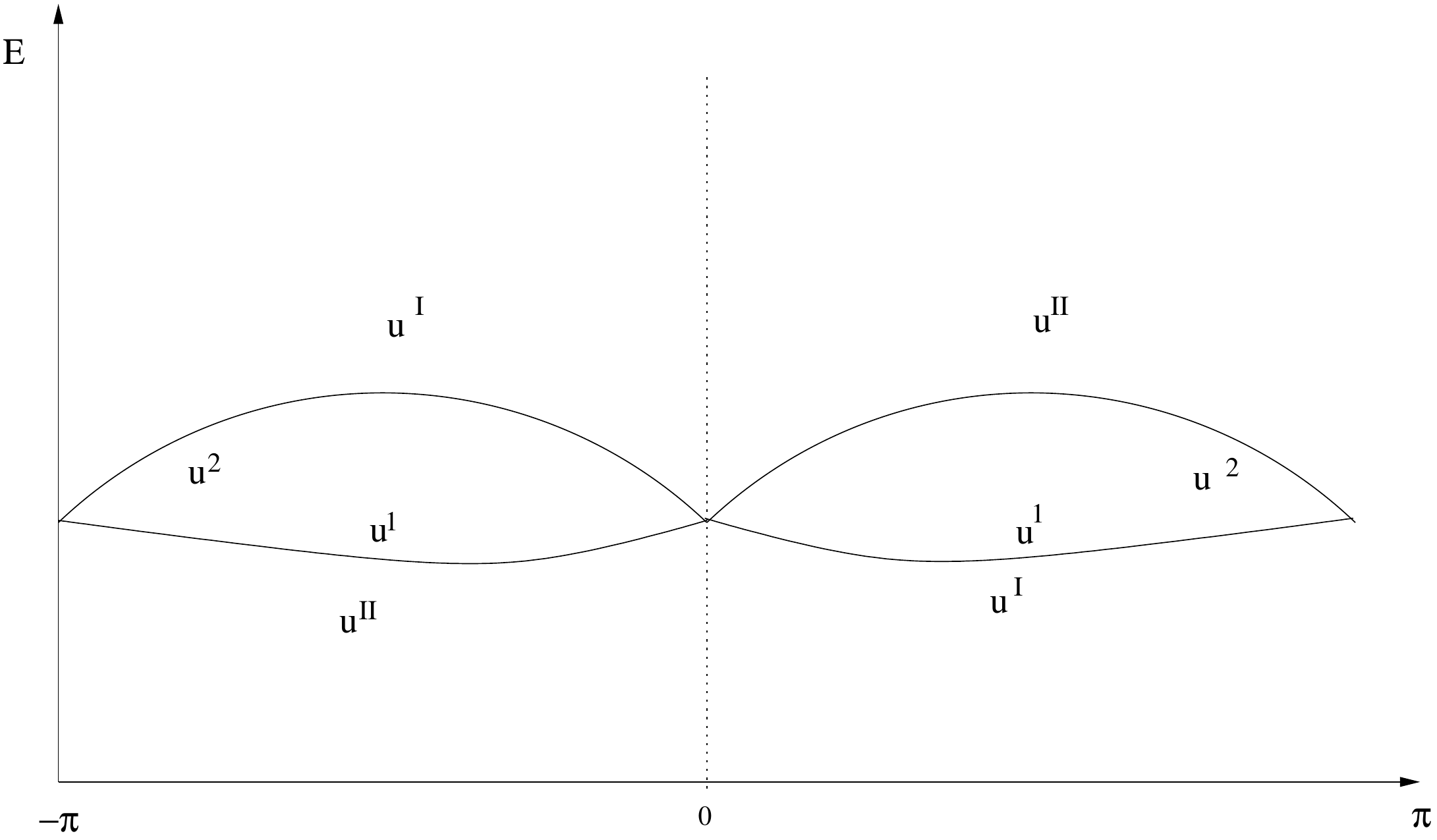}
\end{center}
 \caption{Sections in the Brillouin zone}
\label{fig1}
\end{figure}

Strictly speaking, to follow the literature, we only need the 1--d version
for the base $S^1$, with involution $\T(\bk)=-\bk$ with $\bk\in [0,2\pi]\mod 2\pi$.
In this case we have fixed points $0,\pi$ and $V^+=[0,\pi],V^-=[-\pi,0]$.
We fix continuous normalized sections $u^{1,2}_n(\bk)$ on
$(0,\pi)$ and $(-\pi,0)$ and extend them to $V^{\pm}$. We relabel them with Roman labels $I$ and $II$ as above. This is
illustrated in Figure \ref{fig1}. Then equation (\ref{rel1})
holds and in this basis the matrix of $\Theta$ becomes

\begin{equation}\label{wmatrix}
   w_n(\mathbf{k})  = (\langle u_n^s(-\mathbf{k}) ,  \Theta   u_n^t(\mathbf{k}) \rangle)_{s,t=I,II}
       = \begin{pmatrix}
            0 & -e^{-i\chi_n(\mathbf{k})} \\
            e^{-i\chi_n(\mathbf{-k})} & 0
         \end{pmatrix} \in U(2)
\end{equation}
This matrix is skew--symmetric at fixed points $\bk=-\bk$ and its Pfaffian there, which is just the upper right corner, is $\pf(w_n(\bk))=-e^{-i\chi_n(\mathbf{k})}$.

This type of argument also works for other regions $V=V^+\cup V^-$ with $\T(V^+)=V^-$,
$V_0:=V^+\cap V^- \subset Fix(\T)$ s.t.\ all the path components of $V\setminus V_0 $ are contractible and contained in either $V^+$ or $V^-$, i.e.\ there is no path from $V^+$ to $V^-$ that does not go through a fixed point.

\subsubsection{Definition}

In this subsection, we will briefly recall the definition of the $\mathbb{Z}_2$ invariant $\nu$ using time reversal polarization following \cite{FK06}. Originally a $\mathbb{Z}_2$ invariant 
was introduced by Kane and Mele \cite{KM05} in the study of the quantum spin Hall effect in graphene, whose  geometry is that of the honeycomb model with spin-orbit interaction. 
A simplified account was given in Fu--Kane \cite{FK06}. The Kane--Mele invariant will be interpreted as a mod 2 analytical index, so that it counts the parity of the spectral flow of edge states in $\Z_2$ topological insulators \cite{HK10}.

Given the decomposition of $\Bloch=\bigoplus \V_n$ and
restricting each $\V_n$ over an $S^1$,
the above sections $u_n^{I,II}$ can be used to define
an electric polarization in a cyclic adiabatic evolution. To this end, define the partial polarizations as
the integral of the Berry connections,
\begin{equation*}
   P^s := \int_{-\pi}^\pi a_n^s(k) \,dk,   \quad s=I,II
   \end{equation*}
where the Berry connections  are  locally defined as $a_n^s(k) = i \langle u_n^s(k) | \partial_k | u_n^s(k) \rangle $.
Using the combinations of the partial polarizations, one further defines the charge polarization  $P_\rho $
and the time reversal polarization $P_\theta$ ,
\begin{equation*}
 P_\rho : = P^I + P^{II}, \quad  P_\theta : = P^I - P^{II}
\end{equation*}
For time reversal invariant topological insulators, the charge polarization is always zero, and the time reversal
polarization delivers  the topological $\mathbb{Z}_2$ invariant.

In particular, one computes the time reversal polarization for the
 codimension 1 submanifold  $\{(k,0)|k\in [-\pi, \pi]\}\subset \TT^2$ in 2d. This corresponds to an edge (or edge states) of the effective Brillouin zone, which is a fundamental domain for the time reversal $\Z_2$ action.  
 This embedded $S^1$ has  two fixed points, say $\mathbf{\Gamma} = \{ 0,  \pi \} \subset [-\pi, \pi]$. One computes, see e.g.\ chapter 4 of \cite{S13} :
\begin{equation*}
       P_\theta  = \frac{1}{2 \pi} \int_{-\pi}^\pi dk\, ( a_n^I - a_n^{II})\\
                = \frac{1}{2 \pi i } \left[ \ln \frac{\det (w_n(\pi)) }{\det (w_n(0)) } - 2 \ln  \frac{\pf (w_n(\pi)) }{\pf( w_n(0)) }  \right]
\end{equation*}
Here $w_n$ is the transition matrix \eqref{wmatrix},
After exponentiating both sides and using the identity $e^{\pi i} = -1$, one obtains
\begin{equation*}
  (-1)^{P_\theta} =  \prod_{\Gamma_i = 0, \pi} \frac{\sqrt{\det (w_n(\Gamma_i))} }{\pf (w_n(\Gamma_i))}  =:  \prod_{\Gamma_i = 0, \pi} sgn ({\pf (w_n(\Gamma_i))})
\end{equation*}
It is important to point out that
 $\sqrt{det(w_n(k))}$ is a continuous choice of square root defined on the interval from $0$ to $\pi$. This is possible by extending a choice of a branch of the square root at $0$ along the interval $[0,\pi]$ as in \cite{FK06}. 
 This is what gives the coherent choice of normalization mentioned above, which manifests itself as a choice of the sign of the Pfaffian.
  Since both the Pfaffian and the chosen branch are square roots of the determinant, they differ only by a sign $\pm1$. There is
a more highbrow explanation  in terms of sections which will be presented subsequently.

In terms of the time reversal polarization, the Kane--Mele invariant on $S^1$ relative to a choice of square root of the determinant is defined by
\begin{equation*}
   \nu \equiv  P_\theta \quad (\text{mod 2})
\end{equation*}
In general, the $\mathbb{Z}_2$-valued Kane--Mele invariant $\nu$ is defined by
\begin{equation} \label{disz2}
  (-1)^{\nu} =  \prod_{\Gamma_i \in \mathbf{\Gamma} } \frac{\sqrt{\det w_n(\Gamma_i)} }{\pf (w_n(\Gamma_i))}
\end{equation}
for the fixed points $\mathbf{\Gamma}$ of the time reversal symmetry. Here, again, one has to be careful to say how the branches
of the square root of $det$ are chosen. This is dictated by the embedding of $S^1$s, which pass through the fixed points. In the 2d case these are chosen to be $\{(k,0)\}$ and
$\{(k,\pi)\}$ in \cite{FK06}. This is actually not quite enough as realized in \cite{FK06} and explained in \cite{FKM07}.
Namely one needs a branch of the square root defined on all of $\mathbb{T}^2$. Since $c_1(det \Bloch)=0$ there is indeed a global section. One can then argue that this section
has a square root   \cite{FM13,DG14}, and below \ref{freedpar}.
Notice that this is possible, but it is not possible to find a basis of global sections, since the bundle need not be trivial.
 Unlike the case for $S^1$, for the 2d invariant, the  result is actually invariant under such a choice of section, see \cite{FKM07,FM13}. The 3d case is discussed in the next section.

\subsubsection{Spectral flow}\label{specflow}
In the appendix of \cite{FK06}, Fu and Kane showed that the 2d Kane--Mele invariant is equivalently given by
\begin{equation}\label{obsz2}
  \nu \equiv \frac{1}{2 \pi} \left( \oint_{\partial EBZ}  a - \int_{EBZ} f \right) \quad \text{mod 2}
\end{equation}
where $a$ and $f$ are the Berry connection and curvature, $EBZ$ is the 2d effective Brillouin torus with open cylindrical geometry and its boundary $\partial EBZ$ consists of two circles.
The equivalence can be proved by expressing the partial polarizations by integrals of the Berry connection and curvature.
One has to calculate modulo two since this definition of $\nu$ is not gauge invariant, indeed, the integral of $a$ over the two circles always  changes by an even number under a gauge transformation.

An edge state explanation of the invariant is as follows: the $u^{I,II}$ can be viewed as chiral edge states. At any fixed point these cross each other and, after fixing the Fermi energy to be $E_F =0$,  change signs.
By the time reversal symmetry, it would be enough to consider one chiral edge state, since the other one can be easily recovered by reflection.
The Pfaffian would produce a negative sign  if the spectral flow of the chosen chiral edge state is changed by $+1$ or $-1$
under the gauge transformation  $w_n$.
Thus $\nu$ keeps track of the parity of the change of signs of the Pfaffian under the time reversal symmetry represented by $w_n$, which
can be  interpreted as the evenness or oddness of the spectral flow or a mod 2 analytical index. This is made precise below.

In physical terms, a negative sign of the Pfaffian also corresponds to the existence of an unpaired Majorana zero mode.
Under the bosonization procedure, two Majorana zero modes pair together to create an effective composite boson,
so only the parity of Majorana zero modes matters.
For $\nu \equiv 1$, the effective theory consists of a collection of composite bosons (or none) and an
unpaired Majorana zero mode such as a single Dirac cone.

\subsubsection{Strong $\Z_2$ invariant}

We have defined the Kane--Mele invariant or the analytical $\mathbb{Z}_2$ index above precisely in 2d. One can generalize it to 3d  \cite{FKM07} and higher dimensions using the formula (\ref{disz2}), which is called the strong $\Z_2$ invariant.
But, one can also consider various subtori and their $\Z_2$ invariants, which are called weak $\Z_2$ invariants. This in turn allows one to define the strong invariant rigorously via recursion.
We will go through this in detail for a  3d topological insulator, whose  momentum space (as the bulk space) is assumed to be $\mathbb{T}^3$ with $\T(\bk)=-\bk$.
There are eight time reversal invariant points in total:
$(n_1,n_2,n_3):n_i\in \{0,\pi\}$.

The three standard emdeddings of $\TT^2\to \TT^3$ which have constant $i$-th coordinate equal $0,\pi$ each contain 4 fixed points and give rise to six $\Z_2$ weak invariants $\nu_{n_i=0},\nu_{n_i=\pi}, i=1,2,3$, 
computed by the restriction to the particular $\TT^2$ as a 2d invariant. These 2d  $\mathbb{Z}_2$ invariants are not independent and
the relation between them is given by
\begin{equation*}
   (-1)^{\nu_{n_1 =0}}(-1)^{\nu_{n_1 = \pi}} =  (-1)^{\nu_{n_2 =0}}(-1)^{\nu_{n_2 = \pi}} = (-1)^{ \nu_{n_3 =0}}
   (-1)^{\nu_{n_3} = \pi}=:(-1)^{\nu_0}
\end{equation*}
which collects complete information from eight fixed points for different decompositions of $\mathbb{T}^3$ into the strong invariant $\nu_0$.

In practice, physicists use four $\mathbb{Z}_2$ invariants $(\nu_0; \nu_1\nu_2\nu_3)$ to characterize a 3d topological insulator,
where $\nu_0$ is the 3d $\mathbb{Z}_2$ invariant and $\nu_i \, (i = 1,2,3)$ are 2d $\mathbb{Z}_2$ invariants.

For 3d
physicists often use the momenta $k_i$ to lie in $[0,2]\mod 2$ and the multiplicative form of $\Z_2$ instead of the additive form, i.e.\ $\nu \equiv \pm 1$ instead of $\nu \equiv 0, 1$.
Following these notations   $(\nu_0; \nu_1\nu_2\nu_3)$ are,
\begin{equation}
   \nu_0 := \nu_{n_i =0} \nu_{n_i = 1}, \quad \nu_1 := \nu_{n_1 =1}, \quad \nu_2 :=\nu_{n_2 =1} \quad \nu_3:= \nu_{n_3 =1}
\end{equation}
A product such as above indicates this multiplicative notation.

For instance, a weak topological insulator could have $\nu_1 \equiv 1$,  $\nu_2 \equiv 0$,  $\nu_3 \equiv 1$ but $\nu_0 \equiv 0$,
which is denoted by $(0; 101)$. They are called weak, because they do not have to be stable under general deformations.
We will talk about these four $\mathbb{Z}_2$ invariants again when we discuss the KR-groups of the  3-torus in section \S\ref{Ktheorypar}.

 The 3d Kane--Mele invariant also counts the parity of Dirac cones,
i.e., conical singularities produced by surface states at the fixed points, where
 we regard the codimension--1 sub-manifolds $\mathbb{T}^2$ as  edge spaces of the effective Brillouin zone.
If there exist even number of Dirac cones, then the 3d $\mathbb{Z}_2$ invariant is zero, i.e., $\nu_0 \equiv 0$,
but some 2d $\nu_i\, (i = 1, 2, 3)$ could be non-zero when restricted to a submanifold $\mathbb{T}^2$. In this case,
such time reversal invariant material is called a weak topological insulator.
On the other hand, if there exist odd number of Dirac cones, then
$\nu_0 \equiv 1$ and it is called a strong topological insulator.

\subsubsection{Holonomy}
\label{holpar}
In this subsection, we will express the Kane--Mele invariant (originally in a quotient form) as a difference form with the help of holonomy.

We will consider the Bloch bundle $\Bloch$ restricted to  $R^1$, the effective boundary of the EBZ .
In this region,  there are no  Kramers pairs, since $\Theta$ does not act fiber-wise.
But we can consider either $|u^{II}(\bk)\rangle$ or
$\Theta|u^{I}_n(\mathbf{-k})\rangle$ as interpolating between the quaternionic bases at $0$ and $\pi$, see \S\ref{SympStr}.

Define $\textbf{h}$  as
\begin{equation}
       \mathbf{h} =  \frac{1}{ \pi} \int_0^\pi \, a^{II}(k)-a^{I}(-k)
      \,dk
                 = \frac{1}{ \pi } (\chi_n(\pi) -  \chi_n(0))
\end{equation}
with the help of the relation
$  a_n^{II}({k}) - a_n^{I}(-{k}) = \partial_{k} \chi_n({k})
$ \cite{FK06}.

This can be either seen as a half quaternionic phase winding or a quaternionic holonomy around the loop $-\pi\to \pi$.
This holonomy basically gives the Pfaffian part of the time reversal polarization $P_\theta$ discussed in the previous subsection.

Let us start with $\mathbf{h}$ as the Pfaffian of the time reversal operator $\Theta$ and express the Kane--Mele invariant $\nu$
as the difference of two geometric objects. Since $w_n(\pi)$ and $w_n(0)$ are skew-symmetric matrices,
their Pfaffians are
\begin{equation*}
   \pf(w_n(\pi)) = -e^{-i\chi_n(\mathbf{\pi})}, \quad  \pf(w_n(0)) = -e^{-i\chi_n(0)}
\end{equation*}
Thus
\begin{equation*}
    \ln \pf ({w_n(\pi))} - \ln \pf ({w_n(0))} = 2(k -\ell)\pi i   -i  (\chi_n(\pi) - \chi_n(0))
\end{equation*}
for some integers $k, \ell \in \mathbb{Z}$.
Roughly, the holonomy is the difference of the logarithmic Pfaffians  up to some winding numbers,
\begin{equation*}
    \pi \mathbf{h}   =  i[\ln \pf ({w_n(\pi))} - \ln \pf ({w_n(0))}] + 2 \pi  (k -\ell)
\end{equation*}
To eliminate the ambiguity in choosing the integers from the complex argument of the logarithmic function, we modulo both sides by $2 \pi$,
\begin{equation*}
   {\pi} \mathbf{h}   \equiv  {i}\ln \left[ \frac{\pf (w_n(\pi))}{\pf (w_n(0))} \right] \quad \text{(mod 2$\pi$)}
\end{equation*}
It is easy to get an equality by exponentiating as before,
\begin{equation}
   (-1)^{ \mathbf{h}} =  \frac{\pf (w_n(0))}{\pf (w_n(\pi))}
\end{equation}

The Pfaffian of $w_n$ is  well-defined only at the fixed points $k = 0, \pi$, but the determinant of $w_n$ is defined everywhere for $k \in [0, \pi]$.
Let us start with the equation satisfied by the determinant,
\begin{equation*}
 \frac{1}{2} \int_0^\pi d [\ln \det (w_n(k))]   =  \frac{1}{2}\ln \left[ \frac{\det (w_n(\pi))}{\det (w_n(0))} \right]= \ln \frac{\sqrt{\det (w_n(\pi))}}{\sqrt{\det (w_n(0))}}
\end{equation*}
For any skew-symmetric matrix $A$,  the sign of the Pfaffian satisfies,
\begin{equation*}
   \pf^2A = \det A, \quad sgn(\pf A) = \sqrt{\det A}/\pf A
\end{equation*}
Then the above can be written in the Pfaffian and its sign,
\begin{equation*}
\ln  \frac{\sqrt{\det (w_n(\pi))}}{\sqrt{\det (w_n(0))}}   = \ln \left[ \frac{\pf (w_n(\pi))}{\pf (w_n(0))} \right] + \ln \left[ \frac{sgn(\pf (w_n(\pi)))}{sgn(\pf (w_n(0)))} \right]
\end{equation*}
Change the determinant into a trace, we have
\begin{equation*}
  \frac{1}{2} \int_0^\pi d [\ln \det (w_n(k))]  = \frac{1}{2} \int_0^\pi d \,[tr \ln  w_n(k)] = \frac{1}{2} \int_0^\pi dk \, tr(w_n^{-1}\partial_k w_n)
\end{equation*}
Therefore, we have
\begin{equation*}
   \ln \left[ \frac{\pf (w_n(\pi))}{\pf (w_n(0))} \right] +  \ln \prod_{\alpha = 0, \pi} {sgn(\pf (w_n(\alpha)))} = \frac{1}{2} \int_0^\pi dk \, tr(w_n^{-1}\partial_k w_n)
\end{equation*}

In order to compare it with the holonomy, we exponentiate it,
\begin{equation*}
   \frac{\pf (w_n(\pi))}{\pf (w_n(0))} = (-1)^{\mathbf{n}(w_n)}\prod_{\Gamma_i = 0, \pi} {sgn(\pf (w_n(\Gamma_i)))}
\end{equation*}
where the half winding number $\mathbf{n}$ is defined by
\begin{equation}
   \mathbf{n}(w_n) := \frac{1}{2 \pi i} \int_0^\pi dk \, tr(w_n^{-1}\partial_k w_n)
   \end{equation}
Then the above is equivalent to,
\begin{equation*}
   (-1)^{-\mathbf{h}} = (-1)^{\mathbf{n} + \nu}
\end{equation*}
In other words, the $\mathbb{Z}_2$  invariant $\nu$ is equivalently defined  by
\begin{equation}\label{holz2}
  \nu  \equiv \mathbf{n} - {\mathbf{h}}    \quad \text{(mod 2)}
\end{equation}
In contrast to the Kane--Mele invariant being a quotient in the previous subsection, this difference form is another way to compare the square root of the determinant (as a half winding number)
with the Pfaffian (as a holonomy).

\subsection{Topological band theory} \label{Quatbundle}

With the additional structure given by the time reversal symmetry, the Hilbert bundle modelling the band structure becomes  a Quaternionic vector
bundle. Specifically, the time reversal $\mathbb{Z}_2$ symmetry introduces an involution on the Brillouin torus and a  Quaternionic structure on the physical Hilbert space, that is the sections of the Bloch bundle.
Over the time reversal invariant points, the quaternionic structure gives rise to a symplectic structure of the restricted Bloch bundle.

\subsubsection{Quaternionic structure}\label{SympStr}
In the bundle theoretic framework, one obtains the Hilbert space as the space of sections of the Bloch bundle $\mathcal{H}=\Gamma(\Bloch)$. Since the bundle $\Bloch$ splits as $\bigoplus_n \V_n$, 
so does the Hilbert space as $\bigoplus_n \mathcal{H}_n$, with $\mathcal{H}_n=\Gamma(\V_n)$.

With the extra structure of the time reversal operator $\Theta$,
the  complex Hilbert space $\mathcal{H}(\C)=\mathcal{H}$  can be viewed
as a quaternionic Hilbert space $\mathcal{H}(\mathbb{H})$, by setting $j = \Theta$, with $j^2=\Theta^2=-1$.  The action of $\mathbb{H}=span_{\R} \langle 1,i,j,k\rangle=span_{\C}\langle 1,j\rangle$ is given by  $i$, 
i.e.\ multiplication by $i$, $j$ acting as $\Theta$, which is anti--unitary and hence anti--commutes with $i$,   and $k =ij= i\Theta$.
 $\Theta$ respects the decomposition of $\mathcal{H}$, so that each $\mathcal{H}_n$ becomes a $\mathbb{H}$ vector space $\mathcal{H}_n(\mathbb{H})$ and $\mathcal{H}(\mathbb{H})=\bigoplus_n\mathcal{H}_n(\mathbb{H})$. 
 The fibers of $\mathcal{V}_n$ over a fixed $\bk$ are 2--dimensional complex vector sub--spaces. $\Theta$ in general permutes these sub--spaces, except over the fixed points. At these points the fiber becomes a quaternionic vector space.
 That is $\Bloch|_{\boldsymbol{\Gamma}}$ is a quaternionic bundle and each fiber ${\mathcal{V}_n}_{\Gamma_i} \cong \mathbb{H}$, that is, a  2d $\C$ vector space  becomes a 1d $\mathbb{H}$ vector space. 
 In particular $\mathcal{H}_{\Fix}=\Gamma(\Bloch,\Fix)$ is an $\H$ invariant subspace.  Over $\boldsymbol{\Gamma}$, if $|u_n\rangle$ is a (local) basis over $\H$ then $|u_n\rangle$ and $|\Theta u_n\rangle$ is a basis over $\C$. 
 These are usually put into a so--called Kramers pair
\begin{equation*}
  | \Psi_n \rangle := (| u_n \rangle, \Theta |u_n \rangle) \in \mathcal{H}_{\Fix}(\H)\subset \mathcal{H}_n(\mathbb{H})
  \end{equation*}
then picking $(| u_n \rangle, \Theta |u_n \rangle)$ as a basis, we see that
\begin{equation*}
   \Theta \cdot     \Psi_n =J \Psi_n, \quad  J= i\sigma_y =\begin{pmatrix}
                                   0 & 1 \\
                                   -1& 0
                                 \end{pmatrix}
\end{equation*}
where $\Psi_n$ is viewed as a 2-vector.

The relation to the Pfaffian is as follows:
define a bilinear form on the  complex Hilbert space $\mathcal{H}$ by
\begin{equation}
\label{formdefeq}
   \omega_\Theta(\phi, \psi) = \langle  \Theta \phi | \psi \rangle
\end{equation}
This form becomes a symplectic form for $\V_n$ over the time reversal invariant points; $\omega_\Theta\in \wedge^2(\V_n)^*$.
Thus  $\V_n|_{\Fix} \rightarrow {\Fix}$  is a symplectic vector bundle.
The associated almost complex structure is $J_n=-i\sigma_y$. Hence
the Bloch bundle restricted to the fixed points $\V_n|_{\Fix} \rightarrow {\Fix}$ carries an associated almost complex structure
$J_{\Theta}=\bigoplus_n J_n$. We have that $\Theta=KJ$.

Recall that the compact symplectic group  $Sp(n)$ is defined as the quaternionic unitary group,
\begin{equation*}
   M \in  Sp(n) = U(n, \mathbb{H}), \quad i.e., \quad MM^\dagger = M^\dagger M = 1
\end{equation*}
where $(m_{ij})^\dagger = (\overline{m}_{ji})$ is the combination of taking the matrix transpose and the quaternionic conjugate.
In particular, one has $Sp(1) = SU(2)$.
The Berry connection, which has the local form $a_n(\mathbf{k}) = \langle \Psi_n (\mathbf{k}) |  d  |\Psi_n (\mathbf{k}) \rangle $,
satisfies the identity $a_n + J^{-1} a_n^t J =0$, i.e., $a_n \in \mathfrak{sp}(1)$.

\begin{rmk}
Instead of (\ref{formdefeq})  the matrix coefficients
of $\Theta$ as a morphism $\Hilb\to \bar{\Hilb}$ are
\begin{equation*}
\omega(\phi,\psi)=\langle \phi,\Theta \psi\rangle=\langle \phi |\Theta|\psi\rangle
\end{equation*}
This is actually an anti--linear 2--form and a section of $\wedge^2(\bar\Bloch)^*$. It is related to $\omega_{\Theta}$ by
\begin{equation}
\omega(\phi,\psi)=-\overline{\omega_{\Theta}(\phi,\psi)}
\end{equation}
\end{rmk}

\subsubsection{Quaternionic K-theory}
\label{bundletypepar}
The extra structure that $\Theta$ adds to the bundle has actually been studied by mathematicians and is called a Quaternionic structure
\cite{D69}. In this setup one considers a locally compact topological space $X$ together with an involution $\imath$.
The relevant example here is the time reversal symmetry acting on the Brillouin torus. A Quaternionic bundle on such a space is a bundle $\Bloch$ together with an anti--linear morphism $\chi$, 
which is compatible with $\imath$ and satisfies $\chi^2=-1$. In our case $\chi=\Theta$ and the Quaternionic bundle is simply
 $\BlochT$.

If instead $\chi^2=1$, then the pair is called a Real bundle, which was first considered by Atiyah \cite{A66}. The capital ``Q'' and ``R'' indicate that the
morphism $\chi$ does not necessarily act as a bundle isomorphism, but rather interchanges the fibers over $x$ and $\imath(x)$.
If the involution $\chi$ is trivial, one obtains quaternionic and real bundles.
 Looking at classes of bundles one obtains the respective $K$-theories.
Complex $K=KU$, real $KO$, quaternionic or symplectic $KSp$, Real $KR$ and Quaternionic $KH$ (or $KQ$).
These are alternatively defined using homotopy theory, the Grothendieck construction or stable equivalence \cite{A66}, see below.
As we consider the case with $\Theta^2=-1$, we will be mainly interested in Quaternionic structures. The others appear when there are other symmetries present, see \S\ref{tenpar} below. 
For instance the case with $\Theta^2=1$ represents particles with integer spin \cite{Wigner}.

Denote  $KH(X)$ the K-group of Quaternionic vector bundles over an involutive space $(X, \tau)$.
Sometimes the notation $KQ(X, \tau)$ is also used \cite{DG14}.
If the involution $\tau$ is trivial, then $KH(X)$ is reduced to the quaternionic K-theory $KSp(X) = K_\mathbb{H}(X) = KQ(X, id_X)$.
Similar to KR-theory, the KH-groups have a period 8,
\begin{equation*}
   KH^{i}(X) \simeq KH^{i+8}(X)
   \end{equation*}

The relation between the KH-groups  and  the KR-groups is given by
\begin{equation*}
   KH^{i}(X) \simeq KR^{i-4}(X)
\end{equation*}
Since the Bloch bundle $\BlochT \rightarrow (\mathbb{T}^d, \mathcal{T})$ has
a Quaternionic structure $(\Theta^2 = -1)$ rather than a Real structure,
it is better to use $KH(\mathbb{T}^d)$ for time reversal invariant topological insulators.
However, the above relation tells us that if we  shift the KR-groups by 4,  then the KR-classification \cite{FM13} agrees with the KH-classification \cite{DG14}
for the classification problem of  Bloch bundles over the Brillouin torus for time reversal invariant models.

 \subsubsection{Determinant line bundle}
\label{freedpar}
In this subsection, we first give another equivalent definition of the Kane--Mele invariant by
looking at the determinant line bundle following \cite{FM13}.
Then we discuss the closely related FKMM-invariant in equivariant cohomology following \cite{DG14}.
The calculations of \S \ref{holpar} yield the equivalence to the classical definition of the $\mathbb{Z}_2$ invariant using integrals.

For the rank $2N$ Bloch bundle $\BlochT   \rightarrow \mathbb{T}^d$ consider
its determinant line bundle, which is given by:
$Det\, \Bloch : = \wedge^{2N} \Bloch \rightarrow \mathbb{T}^d$.
Taking the 2N-th exterior power $Det \, {\Theta}$ of ${\Theta}$, we obtain an anti-linear morphism
that squares to $+1$. This yields a Real bundle $(Det \,\Bloch,Det\, \Theta)$,
and when restricted to the fixed point set $\Fix$ a real structure $(Det \,\Bloch|_{\Fix},Det\, \Theta|_{\Fix})$.
The $Det \,\Theta$  invariants of $Det \,\Bloch|_{\Fix}$ form a real sub-bundle, which is denoted  by $\realdet$.

This real bundle has an orientation given by the first component of the Bloch states.
This means that if the $|u_n\rangle$ are a quaternionic basis, then
$|u_1\rangle \wedge \Theta |u_1\rangle \wedge \dots \wedge |u_N\rangle \wedge \Theta| u_N\rangle $
is a canonical
orientation. It is   $Det \, \Theta$ invariant, since
$Det\,\Theta(\phi\wedge\Theta\phi)=\Theta\phi\wedge\Theta^2\phi
=-\Theta\phi\wedge\phi=\phi\wedge\Theta\phi$.
 It is canonical since the space of quaternionic bases is connected.

Alternatively, the symplectic structure $\omega_{\Theta}$ defines an anti--linear 2N form ---aka.\ the Pfaffian form--- which is
\begin{equation*}
   \pf (\omega_\Theta) : = \frac{\omega_\Theta^{\wedge N}}{N!} =\frac{\omega_\Theta\wedge \dots\wedge \omega_\Theta}{N!}
\end{equation*}
It restricts to an $\R$--linear form on the real bundle $\realdet$ yielding an orientation.
This follows from $\omega_{\Theta}(\Theta \, u, \Theta v)=\overline{w_{\Theta}(u,v)}$.

As mentioned above, $c_1(Det \Bloch)=0$ and hence there is a nowhere vanishing section. We can make this section real, i.e.\ land in the
$Det \, \Theta$ invariant part by averaging, see e.g.\
 \cite{FM13} for details.
 We will denote this section by  $s$.
By comparing $s(\Gamma_i)$ with the canonical orientation of the real line $\realdet$,
one defines the $\mathbb{Z}_2$ invariant as the product of signs of the invariant section
\begin{equation*}
   \nu = \prod_{\Gamma_i \in \boldsymbol{\Gamma}}  sgn\, (s(\Gamma_i))  =
   \prod_{\Gamma_i \in \boldsymbol{\Gamma}} \frac{s(\Gamma_i) }{\pf(\Gamma_i)}
\end{equation*}
where $\pf: \boldsymbol{\Gamma} \rightarrow  \realdet$ is a global section representing
 the canonical orientation determined by the Pfaffian form. I.e.\ the sign is positive/negative if $s$ is in the positive/negative half line determined by the orientation.  
 Abusing notation, one can set $s=\sqrt{\det}$ since it is $Det \Theta$ invariant, and rediscover the Kane--Mele invariant. In other words, the $\Z_2$ invariant
  can be defined equivalently by comparing the orientations of the restricted determinant line bundle and the Pfaffian line bundle over the fixed points.

In \cite{DG14}, a so-called FKMM-invariant was constructed almost along the same lines as above.
The restricted determinant line bundle $Det \Bloch|_{\boldsymbol{\Gamma}}$ is trivial
since $\boldsymbol{\Gamma}$ is a finite set. Furthermore there is a canonical isomorphism
\begin{equation*}
   det_{\boldsymbol{\Gamma}}:  Det \, \Bloch|_{\boldsymbol{\Gamma}} \cong {\boldsymbol{\Gamma}} \times \mathbb{C}
\end{equation*}
This section is related to the above equivariant section by $\sqrt{det} = det_{\boldsymbol{\Gamma}}^{-1}$.
The Real determinant line bundle $(Det\, \Bloch, Det \,\Theta)$ has an associated equivariant trivialization
$h_{det}: Det \Bloch \cong \mathbb{T}^d \times \mathbb{C}$,
and its restriction $h_{det}|_{\boldsymbol{\Gamma}}$ gives another trivialization of $Det \, \Bloch|_{\boldsymbol{\Gamma}}$.
The composition
\begin{equation*}
   h_{det}|_{\boldsymbol{\Gamma}} \circ det_{\boldsymbol{\Gamma}}^{-1} : {\boldsymbol{\Gamma}} \times \mathbb{C} \rightarrow {\boldsymbol{\Gamma}} \times \mathbb{C}
\end{equation*}
defines a map
\begin{equation}
   \omega_\Bloch : {\boldsymbol{\Gamma}} \rightarrow \mathbb{Z}_2, \quad \text{s.t.} \quad
   (h_{det}|_{\boldsymbol{\Gamma}} \circ det_{\boldsymbol{\Gamma}}^{-1})(x, \lambda) = (x, \omega_\Bloch(x) \lambda )
\end{equation}
Then the FKMM-invariant is defined as the equivariant homotopy class of $\omega_\Bloch$ up to  equivariant gauge transformations
\begin{equation}
   \kappa(\Bloch) := [\omega_\Bloch] \in [\boldsymbol{\Gamma}, U(1)]_{\mathbb{Z}_2} / [\mathbb{T}^d, U(1)]_{\mathbb{Z}_2}
\end{equation}
In general, $\kappa$ defines an injective map from Quaternionic vector bundles to relative equivariant Borel cohomology classes
over an involutive space $(X, \varsigma)$,
\begin{equation}
   \kappa: Vect_\mathcal{Q}^{2m}(X, \varsigma) \rightarrow H^2_{\mathbb{Z}_2}(X, X^\varsigma; \mathbb{Z}(1))
\end{equation}
See \S 4 of \cite{DG14} for more details about the injectivity, the equivalence between the FKMM-invariant and the Kane--Mele invariant
was also proved in that section.

\subsection{Homotopy theory}
\label{homotopypar}

The study of homotopy theory of time reversal invariant  Hamiltonians can be traced back to  the 80's  \cite{ASSS88, ASS83}.
Denote the space of non--degenerate Hamiltonians, i.e.\  non--degenerate $k \times k$ Hermitian matrices
 by $\mathfrak{H}_k$.
  It is homotopic to
$U(k)/ U(1)^k$. The lower homotopy groups were computed in \cite{ASS83},
\begin{equation*}
   \pi_1(\mathfrak{H}_k) = 0, \quad \pi_2(\mathfrak{H}_k) = \mathbb{Z}^{k-1}
\end{equation*}
In particular, for the 2d case (i.e., $k =2$) $ \pi_2(\mathfrak{H}_2)  = \mathbb{Z}$, which reflects the existence of the TKNN integers (or the first Chern number) as
the $\Z$--invariant.

As mentioned previously, from  \cite{vNW}, we know that for a family of  Hamiltonians $H \in \mathfrak{H}$, its eigenvalue degeneracy is generically
codimension three, so in 3d a $S^2$ can be used to enclose the monopole and the first Chern number of the restriction of the Bloch bundle provides  a topological invariant, see \S\ref{Diracsec}.
This corresponds to the fact $\pi_2(\mathfrak{H}_2)=\Z$.

If one includes time reversal symmetry, two more spaces become of interest.
The first is $\CC$, which is the subspace of Hermitian matrices allowing possible pair degeneracies, that is Eigenvalues 1 and 2 or Eigenvalues 3 and 4 may be degenerate and so forth. 
This is the space for families of Hamiltonians which have at most Kramers degeneracy.
The second is the subspace $\CQ$
\begin{equation}
  \CQ = \{ H \,|\, \Theta H\Theta^{-1} = H \}
\end{equation}
where now $\Theta$ is the given action of time reversal on the Hilbert space, that is an anti--unitary operator with $\Theta^2=-1$. To not overburden the notation  we omitted the index $k$, 
but we will always assume that the dimension of the family of Hamiltonians is fixed. Moreover, for time reversal invariant systems the interesting case is when the Bloch bundle has even rank, 
so we will assume that $k=2n$. In this case, independent of the particular choice $\Theta$, each element in $\CQ$ can be represented by a non-degenerate quaternionic $n \times n$ Hermitian matrix, and the space $
\CQ$ is homotopic to $Sp(n)/ Sp(1)^{ n}$, see e.g.\ \cite{ASSS88}.

The relation is that the family of Hamiltonians over the Brillouin torus can be viewed as a map $H:\TT^d\to \mathfrak{H}$. If we add the time reversal symmetry, 
we additionally have the two $\Z_2$ actions $\T$ on $\TT^d$ and conjugation by $\Theta$ on $\mathfrak{H}$. The map $H$ then becomes a $\Z_2$ equivariant map.
The condition that there are only the degeneracies forced by Kramers degeneracy implies that in fact $H:\TT^d\to \CC$. Since this map is still equivariant it is
determined by the restriction $H|_{EBZ}:EBZ\to \CC$. Said differently,
under the time reversal symmetry, we only consider the half  Brillouin torus giving a fundamental domain for the $\Z_2$ action, i.e. the EBZ, since the other half can be
easily recovered by reflection.
Furthermore, at the fixed points $\Gamma_i\in \boldsymbol{\Gamma}$ the Hamiltonians are invariant, so that $H|_{\boldsymbol{\Gamma}}\to \CQ\subset \mathfrak{H}$.

The lower
homotopy groups of time reversal invariant Dirac Hamiltonians are known, see \cite{ASSS88}:
\begin{equation*}
   \pi_1(\CQ_n) = \pi_2(\CQ_n) = \pi_3(\CQ_n) = 0, \quad \pi_4(\CQ_n) \simeq \mathbb{Z}^{n-1}
\end{equation*}

Analogously to the above, for a generic time reversal invariant
 family of Dirac Hamiltonians $H \in \CQ_2$, its eigenvalue degeneracy has codimension five and Kramers doublets are distinguishable in 4d
families,  e.g.\ by enclosing the degeneracies by $S^4$.
In other words, the homotopy classes of maps $[S^4\, , \, \CQ_2] =\pi_4(\CQ_2)=\Z$ are labeled by $\Z$, which is just
 the second Chern number $c_2$ of the Bloch bundle. Thus $c_2$ manifests itself
as a fundamental invariant for time reversal invariant models, which follows from $\pi_4(\CQ)=\Z$.
As before this yields no direct invariants in dimensions less than 4. For other roles of $S^d$, see \S\ref{EBZpar}.

In \cite{MB07}, Moore and Balents proposed that one could cap the two ends of the cylinder $C=S^1\times I$,  representing the effective Brillouin zone to obtain a topological invariant. 
This capping turns $C$ into an $S^2$ and one can then regard an induced family of Hamiltonians on $S^2$, which however depends on the chosen extension to the caps. By a homotopy argument
 which will be explained below, they
obtained the topological $\mathbb{Z}_2$ invariant.

First,  Moore and Balents assert that $\pi_1(\CC) = 0$, which implies that the map $H$ restricted to any of the two circles in the boundary of the effective Brillouin zone
to $\CQ$ is homotopic to the constant map. It is claimed in \cite{MB07}
that this homotopy can be made into a homotopy with two additional properties, which is what they define to be a contraction. The first is that the homotopy is actually a homotopy of equivariant maps, and secondly,
that homotopy $f(\theta,\lambda):S^1\times I\to \CC$ satisfies that $f(\theta,\pi)\in \CQ$.
This can be made rigorous by considering relative homotopy groups $\pi_i(\CQ,\CC,H_0)$ and the corresponding long exact sequence, see e.g. \S4.1 in \cite{H}: Consider $H|_{S^1}\to \CC$ where $S^1$ is one of the boundaries of the EBZ cylinder. 
This map is actually still $\Z_2$ equivariant, so that it is fixed by its restriction to the upper half circle, which is an interval. The restriction to this interval is a path $p$ from $H_0\in \CQ$ to a point $H_\pi\in \CQ$ and 
hence in $\pi_1(\CQ,\CC,H_0)$. From the long exact sequence of relative homotopy groups this group is trivial and hence the path can be homotoped through paths $p_t$ to a path $p_1$ entirely inside $\CQ$.
Considering the loops $l_t=p_t$ for points in the upper half--circle and
$l_t=\Theta p_t\Theta^{-1}$ for the lower half--circle, we obtain an equivariant deformation of $H|_{S^1}$ to a map $l_1:S^1\to \CQ$, with $l_t(\pi)\in \CQ$. Since $\pi_1(\CQ)=0$,
we can further contract to the constant loop, again in an equivariant fashion,
obtaining, finally, a contraction in the above sense.

Hence the map from the cylinder representing the effective Brillouin zone can be replaced by a map from the 2-sphere, obtained by contracting the boundary circles to a point. More precisely, any map $H:EBZ\to \CC$  
is homotopic to a map that  factors through the quotient space
$S^2=EBZ/\sim$ where $\sim$ is the equivalence relation identifying all points in each boundary circle.  This can also be visualized by gluing two caps onto the cylinder. The resulting map from $\tilde{H}:S^2\to \CC$
produces a first Chern number, which will be denoted by $c_1(\tilde{H})$. This first Chern number depends on the choice of homotopy, though. What is argued in \cite{MB07}, is that choosing different homotopies changes 
the Chern number by an even integer: $c_1(\tilde{H})\equiv c_1(\tilde{H}') \mod 2$. The argument relies on the fact that homotopies can be chosen to be contractions and that the lower homotopy groups $\pi_1,\pi_2$ of $\CQ$ vanish.

After eliminating the ambiguity of extra even Chern numbers from choosing contractions,
a topological $\mathbb{Z}_2$ invariant is obtained by capping the cylinder.
This procedure to eliminate the ambiguity by a homotopy argument is in line with the solution to the gauge ambiguity in \eqref{obsz2}.

\section{$\Z_2$ invariant as an index theorem}
\label{indexpar}
In this section, we show how the $\Z_2$ invariant can be
viewed both as an analytical index based on the Maslov index
and as a topological index by the Chern--Simons invariant.
 We have explained that   the Kane--Mele invariant counts the parity of the spectral flow of edge states in \S \ref{specflow} and that this in turn, in the language of index theory, is the analytical $\mathbb{Z}_2$
index. The topological $\mathbb{Z}_2$ index is the focus of the second subsection.   In physics, it corresponds to  the  Chern--Simons invariant, more precisely, the mod 2 version of the topological term
 in the Wess--Zumino--Witten (WZW) model. The last subsection is a summary of the relations between the different interpretations of the topological $\mathbb{Z}_2$ invariant.

\subsection{Analytical $\mathbb{Z}_2$ index}
\label{Z2invsec}
 The mod $2$ analytical index was first introduced by Atiyah and Singer \cite{AS69, AS71} for real skew-adjoint elliptic operators $P$,
\begin{equation} \label{Amod2}
 ind_2 (P) \equiv \dim \ker (P) \quad \text{(mod 2)}
 \end{equation}
which can be equivalently described by a mod 2 spectral flow.  We argue that the Kane--Mele invariant can be interpreted as a mod $2$ analytical index of the effective Hamiltonian of Majorana zero modes, 
which is explained clearly in our work \cite{KLW15}. For the original Kane-Mele invariant, a possible connection was  first observed by Tony Pantev, see  reference [14] in \cite{KM05}.

In the quantum spin Hall system, the Kane--Mele invariant $\nu \equiv 0$ is equivalent to the fact that
the Fermi level intersects a chiral edge state even times, while $\nu \equiv 1$ means the intersection number
is odd. This is exactly the definition of the mod 2 analytical  index  \eqref{Amod2}. Here we
put it into the symplectic setting and  give an equivalent definition of the Kane--Mele invariant using the Maslov index as in \cite{ASV13}, which gives a geometric realization of the analytical index.

\subsubsection{Maslov index}

Assume that the Fermi level sits inside the band gap between the conduction and valence bands,
in this subsection we view the Fermi level as a fixed reference Lagrangian
subspace $\mathbb{R} \subset \mathbb{R}^2 $. From the helical edge state, we pick one chiral edge state and model it by
a continuous piecewise-smooth real-valued function, which is also modeled by one Lagrangian subspace or several Lagrangians in $\mathbb{R}^2$. If we count
the intersection points of the chosen chiral edge state with the Fermi level, then the 2d $\mathbb{Z}_2$ invariant can be equivalently defined by
the Lagrangian intersection number or Maslov index as an edge $\mathbb{Z}_2$ index \cite{ASV13}.

It is well known that the Maslov index can be characterized by a set of axioms, which has different geometric
and analytical realizations
such as eta invariant or spectral flow.
In this subsection, we recall the relations between the Maslov index, eta invariant and spectral
flow within $\mathbb{R}^2$  following \cite{CLM94}.
More generally, on manifolds with boundary the spectral flow of a family of Dirac operators
can also be computed by the Maslov triple index, the interested reader may consult for example \cite{KL04}.

In the simplest symplectic vector space $(\mathbb{R}^2, \omega)$ with the standard symplectic form $\omega$,
 $L \subset \mathbb{R}^2$ is a Lagrangian subspace, or simply Lagrangian, of $(\mathbb{R}^2, \omega)$ if $L = L^\perp$,
which is equivalent to $L$ being 1-dimensional in $\mathbb{R}^2$. Denote the 1-dimensional Lagrangian Grassmannian $Gr(1, 2)$ in $(\mathbb{R}^2, \omega)$ by
$\Lambda(1)$, and each element in $\Lambda(1)$ is represented by a line $L(\theta)$ with slope $\theta$. The square function
\begin{equation*}
   Sq: \Lambda(1) \rightarrow S^1; \quad L(\theta) \mapsto e^{2 i \theta}
\end{equation*} induces a diffeomorphism
$   \Lambda(1) \simeq U(1)/O(1) \simeq \mathbb{RP}^1$
and the fundamental group of $\Lambda(1)$ is
$   \pi_1(\Lambda(1)) = \pi_1(S^1) = \mathbb{Z}$.

In $(\mathbb{R}^2, \omega)$, the geometric Maslov index is defined as the intersection number of a line $L(\theta) \in \Lambda(1)$ with
a reference line usually fixed as $\mathbb{R} = L(0)$. Remember that in the 2d quantum spin Hall system,
we model the chiral edge states by 1d Lagrangians, and we are interested in their intersections with the Fermi level.
One version of  analytical Maslov index is defined by the eta invariant of a real self-adjoint Dirac operator
$D(L_1, L_2)$ where $L_1, L_2 \in \Lambda(1)$. Another version of analytical Maslov index is defined by the spectral flow of
a family of Dirac operators
$\{ D(L_1(t), L_2(t)), a \leq t \leq b \}$. In other words, the geometric Maslov index can be interpreted
as the spectral flow of a family of Dirac operators, which is intimately related to the eta invariant.

We now show how to compute the analytical Maslov indices explicitly in $(\mathbb{R}^2, \omega)$.
Denote the unit interval by  $I=[0,1]$, let $W^{1,2}(I;L_1, L_2)$ be the Sobolev space as the completion of the space of
 paths connecting two Lagrangians $L_1, L_2$, i.e.
\begin{equation*}
   \phi \in C^1(I, \mathbb{R}^{2}) \quad s.t. \quad \phi(0) \in L_1, \phi(1) \in L_2
\end{equation*}
The Dirac operator $D(L_1, L_2)$ defined by
\begin{equation}
   D(L_1, L_2): W^{1,2}(I;L_1, L_2) \rightarrow L^2(I, \mathbb{R}^{2}); \quad \phi \mapsto -i \frac{d \phi}{dt}
\end{equation}
is a Fredholm operator and its kernel is given by constant functions
\begin{equation*}
  Ker(D(L_1, L_2)) = \{ \phi(t) = constant\,\, point \in L_1 \cap L_2 \}
\end{equation*}

When $Re(s) \gg 0$, the $\eta$-function of a self-adjoint elliptic operator $D$ is defined by
\begin{equation}
  \eta(D, s) = \sum_{\lambda> 0} \frac{1}{\lambda^s} - \sum_{\lambda <0} \frac{1}{(-\lambda)^s}
\end{equation}
which admits a  meromorphic extension to the complex plane and is regular at $s=0$. One defines the eta invariant
as $\eta(D) := \eta(D, 0)$, and the reduced eta invariant $\tilde{\eta}(D)$ as
\begin{equation}
    \tilde{\eta}(D) := \frac{\eta(D)  + \dim (\ker D)}{2}
\end{equation}
Roughly, the eta invariant is  the number of positive eigenvalues minus the number of negative eigenvalues.
In symplectic geometry, one defines the analytical Maslov index  by the eta invariant
\begin{equation}
   \mu_{a} := \eta(D(L_1, L_2))
\end{equation}
We stress that this analytical Maslov index is real-valued.

For example,  consider two typical Lagrangians $\mathbb{R}$ and $L(\theta) = \mathbb{R}\{e^{i\theta}\}$ in $(\mathbb{R}^{2}, \omega)$,
the path connecting them can be taken to be
\begin{equation*}
   \phi(t) = e^{i(\ell \pi  + \theta)t}, \quad \text{for some} \quad \ell \in \mathbb{Z}
\end{equation*}
where the integer $\ell$ is added since the Lagrangian $\mathbb{R}\{e^{i\theta} \}$ is invariant by multiplying a minus sign. Since
\begin{equation*}
   D(\mathbb{R}, L(\theta)) \phi = -i \frac{d \phi}{dt} = (\ell \pi + \theta) \phi
\end{equation*}
the eta function of $ D(\mathbb{R}, L(\theta))$ is
\begin{equation*}
   \eta(D(\mathbb{R}, L(\theta)), s) = \frac{1}{\pi^s} \left[ \zeta_H(s, \theta/\pi) - \zeta_H(s, 1- \theta/\pi) \right]
   \end{equation*}
where $\zeta_H(s, x ) $ is the Hurwitz zeta function $   \zeta_H(s, x ) = \sum_{n=0}^\infty (n+ x)^{-s} $
for $0< x < 1$.  The Hurwitz zeta function
has a  regular point at $s=0$, indeed
$    \zeta_H(0, x) = \frac{1}{2} - x$.
Hence the real-valued Maslov index is easily obtained by direct computation
\begin{equation}
   \mu_a = \eta( D(\mathbb{R}, L(\theta)), 0) = 1- \frac{2\theta}{\pi}
\end{equation}

In order to be compatible with the integral Lagrangian intersection number, i.e., the geometric Maslov index, the integer-valued
analytical Maslov index is defined as follows.
Define the canonical 1-form  on $\Lambda(1)$ by pulling back the standard 1-form on $S^1$, %(Keller-Arnold-Maslov form)
\begin{equation*}
   \Omega :=(Sq)^* \left( \frac{d\theta}{2 \pi} \right)
\end{equation*}
which is the generator of the cohomology group
$   H^1(\Lambda(1), \mathbb{Z}) \simeq H^1(S^1, \mathbb{Z}) \simeq \mathbb{Z}$.
Now the integer-valued analytical
   Maslov index is defined for a pair of continuous and piecewise smooth Lagrangians $f(t) = (L_1(t), L_2(t))$, for some interval $a \leq t \leq b$,
\begin{equation}
   \mu_a(f) := \int_a^b [ L_2^* \Omega  - L_1^*\Omega ] + \frac{1}{2} \tilde{\eta}(D(L_1, L_2))|_a^b
\end{equation}
where
\begin{equation*}
   \int_a^b L_s^*\Omega = \frac{1}{\pi} \int_a^b \sum_{i =1}^n d\theta_{si} = \frac{1}{\pi} \sum_{i =1}^n [\theta_{si}(b) -  \theta_{si}(a) ], \quad s = 1,2
\end{equation*}

In $(\mathbb{R}^2, \omega)$, let $L_1 = \mathbb{R}, L_2 = \mathbb{R} \{ e^{i\theta}\}$, then it is easy to compute the eta invariant
\begin{equation*}
    \eta(D(L_1, L_2)) = \left\{
  \begin{array}{l l}
    -1 - \frac{2 \theta}{\pi} & \quad  \text{for }  -\frac{\pi}{4} \leq \theta < 0 \\
    1 - \frac{2\theta}{\pi} & \quad  \text{for } 0< \theta \leq \frac{\pi}{4} \\
    0 & \quad \text{for } \theta= 0

  \end{array} \right.
\end{equation*}
 the pull-backs are given by $   L_1^*\Omega = 0, \,\, L_2^*\Omega = {d\theta}/{\pi}$.
For $f(t) = (\mathbb{R},\mathbb{R} e^{it})$, it follows that
\begin{equation*}
   \mu_a(f|[-\frac{\pi}{4}, 0]) = 0, \quad \mu_a(f|[0,\frac{\pi}{4}]) = \mu_a(f|[- \frac{\pi}{4}, \frac{\pi}{4}]) = 1
\end{equation*}

The spectral flow of a family of self-adjoint elliptic operators, in practice Dirac operators, $D_t = D(L_1(t), L_2(t)), a \leq t \leq b$
is defined as usual by the net number of eigenvalues changing sign ($+1$ for increasing, $-1$ for decreasing)
from $a$ to $b$, denoted by $sf(D_t)$. A second integer-valued analytical  Maslov index is defined by the
spectral flow of the family of Dirac operators associated with the pair $f(t) = (L_1(t), L_2(t))$,
\begin{equation}
   \mu_a(f) := sf(D(L_1(t), L_2(t)) ), \quad a \leq t \leq b
\end{equation}

Recall that $D(\mathbb{R}, \mathbb{R} e^{i\theta})$ has eigenvalues
$\pi \ell + \theta$ for $-\frac{\pi}{4} \leq \theta \leq \frac{\pi}{4}$.
When $\theta$ approaches $0$ from the left,  the eta invariant has the limit
\begin{equation*}
    \lim_{\theta \rightarrow 0-} \eta(D(L_1, L_2)) =  \lim_{\theta \rightarrow 0-} \left( -1  - \frac{2 \theta}{\pi} \right) = -1
\end{equation*}
Meanwhile, we also have the right limit
\begin{equation*}
    \lim_{\theta \rightarrow 0+} \eta(D(L_1, L_2)) =  \lim_{\theta \rightarrow 0-} \left( 1  - \frac{2 \theta}{\pi} \right) = 1
\end{equation*}
Immediately, the spectral flow is easily obtained from the jump of the eta invariant crossing $0$, i.e.
\begin{equation*}
   sf(D(L_1(t), L_2(t)) ) = \frac{1}{2} [\eta(D(L_1, L_2), 0+) -  \eta(D(L_1, L_2), 0-)] = 1
\end{equation*}
The analytical Maslov index unifies the spectral flow and the eta invariant in the symplectic setting, that is,
the spectral flow of a family of Dirac operators can be computed as the variation of eta invariants.

To end this subsection, we define  the  edge  $\mathbb{Z}_2$ index  \cite{ASV13}
by the Maslov index $\mu$ modulo two,
\begin{equation}
  \nu \equiv \sum_i \mu (\mathbb{R}, L(\theta_i)) \quad  \text{(mod 2)}
\end{equation}
where the Fermi level is fixed as $\mathbb{R} = L(0)$ and one chiral edge state is assumed to be piecewisely modeled by
a finite collection of 1d Lagrangian subspaces $L(\theta_i) \in \Lambda(1)$.
The equivalence with the  Kane--Mele invariant can be seen by expressing the edge  $\mathbb{Z}_2$ index as
the winding number of a closed unitary path, for more details see \cite{ASV13}.

\subsection{Topological $\mathbb{Z}_2$ index}

In this subsection, we will discuss how to compute the $\mathbb{Z}_2$ invariant of a 3d topological insulator by the topological $\mathbb{Z}_2$ index.
One starts out with a 4d time reversal invariant model and applies dimensional reduction to obtain a 3d Chern--Simons field theory.
From the 3d bulk Chern--Simons (CS) theory, one obtains a 2d topological Wess--Zumino--Witten (WZW)  term by the CS/WZW duality.
It is shown in \cite{FM13, KLW15, WQZ10} that for a 3d topological insulator, the Chern--Simons invariant, i.e., the WZW term, and the Kane--Mele invariant
are equivalent, and
for this reason the topological $\mathbb{Z}_2$ invariant  will be called the Chern-Simons invariant in this subsection. In string theory,  the topological WZW term naturally appears in the bosonization process \cite{W84},
and its  global $SU(2)$ anomaly  explains why it is a mod 2 index \cite{W82}.

Mathematically,  the Chern--Simons invariant is computed by the odd topological index, that is, the integral of the odd Chern character of a specific gauge transformation, 
which is an application of the odd index theorem for Toeplitz operators on compact manifolds.
So the topological $\Z_2$ invariant is identified as  the mod 2 reduction of the odd index theorem in 3d.
Or equivalently,  the $\mathbb{Z}_2$ invariant is the mod 2 version of the Witten index $Tr(-1)^F$,
where  $F$ counts  the number of Majorana zero modes such as Dirac cones. In a modern language, the
topological $\mathbb{Z}_2$ index is obtained by pairing the K-homology of Majorana zero modes with the K-theory of the Bloch bundle, for details see \cite{KLW15}.

\subsubsection{CS/WZW correspondence} \label{CWcorr}

Zhang et al. \cite{QHZ08} first considered a time reversal invariant (4+1)-dimensional model generalizing the
(2+1) quantum Hall system, the effective action functional was obtained by linear response theory,
\begin{equation}
\label{4dactioneq}
   S_{4d} = \frac{c_2}{24\pi^2} \int d^4x dt \, \epsilon^{\mu\nu\rho\sigma\tau} A_\mu\partial_\nu A_\rho \partial_\sigma A_\tau
\end{equation}
where $c_2$ is the second Chern number
\begin{equation}
\label{c2eq}
   c_2(f) = \frac{1}{32\pi^2} \int d^4k \, \epsilon^{ijkl}tr( f_{ij}f_{kl} )
\end{equation}
with non-abelian local Berry  connection and curvature defined as before,
\begin{equation*}
  a_i = i \langle \psi | \partial_i | \psi \rangle, \quad   f_{ij} = \partial_ia_j - \partial_j a_i + i [a_i, a_j]
  \end{equation*}

In order to get the  effective action functional of a 3d topological insulator,
 dimensional reduction  was applied to the above effective action $S_{4d}$ and the second Chern number as well  \cite{QHZ08}.
The idea of dimensional reduction is to replace one spatial or momental dimension by an adiabatic parameter,
and then integrate out this adiabatic parameter carefully.

 For the external gauge field,
one can reduce the dimension by fixing a convenient gauge such as the Landau gauge.
As for the 2nd Chern number $c_2(f)$,  one  considers the Chern--Simons class because of its relation with the second Chern character
on the level of characteristic classes.
Indeed, the Chern--Simons form $cs_3(a, f)$
is the boundary term of  the 2nd Chern character, for example see \cite{N03},
\begin{equation*}
   d \,cs_3(a, f) = ch_2(f)
\end{equation*}
In addition, the Chern character can be expressed in Chern classes, in our case
\begin{equation*}
   ch_2(f) = \frac{1}{2}[c_1^2(f) - 2c_2(f)] = -c_2(f)
\end{equation*}
since  $c_1(f) =0$ for time reversal invariant models.

After dimensional reduction,  the  effective action for a 3d topological insulator is given by
\begin{equation}
  S_{3d} = \frac{P_3}{16\pi^2} \int d^3x dt \, \epsilon^{\mu\nu\rho\sigma}F_{\mu\nu}F_{\rho\sigma}
\end{equation}
where the magneto-electric polarization $P_3$ is defined as the Chern--Simons action of the Berry connection,
\begin{equation*}
    P_3 (a_i)= \frac{1}{16\pi^2} \int d^3k \, \epsilon^{ijk} tr  (a_if_{jk}  - \frac{1}{3} a_i [a_j, a_k])
\end{equation*}
It is related to the $\theta$-parameter known in condensed matter physics by
\begin{equation}
\label{thetaeq}
   \theta  := 2\pi P_3 \in \{0,\pi \} \mod 2\pi
\end{equation}

In general, the Chern--Simons action is gauge invariant up to a winding number. More precisely,
under a gauge transformation,
\begin{equation*}
  a_i \mapsto a_i^g = g^{-1} a_i g - g^{-1} d_i g, \quad \text{for} \,\, g: \mathbb{T}^3 \rightarrow U(n)
\end{equation*}
one has
\begin{equation}
  \Delta P_3 = P_3(a_i^g) - P_3(a_i) = \frac{1}{24 \pi^2} \int d^3k \, \epsilon^{ijk} tr(g^{-1} d_i g g^{-1} d_j g g^{-1} d_k g)
\end{equation}
the right hand side is an integer, namely the winding number of $g$. It is also known as the topological WZW term.

The Chern--Simons invariant $ \upsilon$  of a time reversal invariant system is defined in \cite{QHZ08} by
the change of magneto-electric polarization under the gauge transformation induced by the time reversal symmetry modulo two.
More precisely,  if  $w$ is the specific gauge transformation induced by the time reversal symmetry
 given  by the transition matrix  \eqref{wmatrix},  the Chern--Simons invariant is defined by
\begin{equation} \label{intz2}
     \upsilon \equiv \frac{1}{24 \pi^2} \int_{\mathbb{T}^3} d^3k \,  tr(w^{-1} dw)^3 \quad (\text{mod 2})
\end{equation}

Now the Bloch bundle splits as before, so that we have a block form of $w\in U(2)\times \cdots \times U(2)\subset U(2N)$. Furthermore,
by the property of the transition matrix, the gauge group can be reduced from $U(2)$ to $SU(2)$, see \cite{WQZ10},
hence  $ \upsilon$ is the mod 2 version of the topological term in the $SU(2)$ WZW model.
So the chiral edge states can be modeled by some current algebra, which is a Kac--Moody algebra in conformal field theory, the interested reader may consult \cite{W84}.

Up to now, we considered the 3d Chern-Simons theory and the associated topological WZW term.
A similar argument can be applied to the 1d Chern-Simons theory, which can be viewed as a boundary theory of a 2d bulk theory.
The time reversal symmetry also induces a specific gauge transformation, which can be plugged into the 1d WZW term,
i.e., the winding number.  The mod 2 version of this 1d
WZW model can be used to study the continuous model of time reversal invariant Majorana chains.

The physical picture behind the topological WZW term is the bosonization process, that is, two Majorana  zero modes, such as two Dirac cones,
pair together to
become an effective composite boson. The topological WZW term as an action functional describes a  bosonic theory equivalent
to the fermionic theory of Majorana zero modes. The built-in global SU(2) anomaly of the WZW term explains why we have to modulo two, i.e.,
the parity anomaly, which will be discussed in a later subsection.

To sum it up, the CS/WZW duality gives the bulk-edge correspondence in 3d topological insulators. More precisely,
the Chern--Simons action  of the Berry connection describes a 3d
topological insulator in the bulk, while the topological WZW term characterizing the edge states is the action  of
the specific gauge transformation induced by the time reversal symmetry.
 In other words, the bulk field theory is the 3d Chern--Simons
field theory, and the boundary field theory is the 2d Wess--Zumino--Witten model.

\subsubsection{Odd Chern character} \label{oddCh}
From the last subsection, we know that the Chern--Simons invariant can be defined by the mod 2 WZW term of a specific gauge transformation.
It is well-known that the WZW term gives rise to a topological index of Toeplitz operators.
In this subsection, we will review the odd Chern character of gauge transformations and its spectral flow  following \cite{G93},
so that the Chern--Simons invariant
will be naturally interpreted as the mod 2 spectral flow through odd Chern character.

For two connections $\mathcal{A}_0$ and $\mathcal{A}_1$ on some vector bundle, the relative Chern--Simons form is defined by \cite{G93}
\begin{equation}
   cs( \mathcal{A}_0, \mathcal{A}_1) : = \int_0^1 tr(\dot{\mathcal{A}}_t e^{\mathcal{A}_t^2}) dt
\end{equation}
where
\begin{equation*}
   \mathcal{A}_t = (1-t) \mathcal{A}_0 + t \mathcal{A}_1, \quad \dot{ \mathcal{A}_t} =  \mathcal{A}_1 - \mathcal{A}_0
\end{equation*}
There exists a transgression formula connecting the Chern--Simons form and Chern characters,
\begin{equation}
   d\, cs( \mathcal{A}_0, \mathcal{A}_1) = ch( \mathcal{A}_1 ) - ch( \mathcal{A}_0 )
\end{equation}
where the Chern character is defined as usual, $ch(\mathcal{A}) = tr(e^{\mathcal{A}^2})$.

Furthermore, given a  map from a compact $(2k+1)$-dimensional spin manifold  $M$  to the unitary group $g: M \rightarrow U(n)$,
its homotopy class is an element in the odd $K$-group, i.e., $ [g] \in K^{-1}(M)$. The odd Chern character of $g$ is defined by \cite{G93}
\begin{equation}
   ch(g) := \sum_{k=0}^{\infty} (-1)^k \frac{k!}{(2k+1)!} tr[(g^{-1}dg)^{2k+1}]
\end{equation}
which is a closed form of odd degree. Or equivalently, $ch(g)$ is the relative Chern-Simons form $cs(d, g^{-1}dg)$ by Taylor expansion.
As a classical example, the degree of $g: S^{2k+1} \rightarrow U(n)$ is given by
\begin{equation*}
   \deg(g )=  - \left( \frac{i}{2\pi }\right)^k \int_{S^{2k+1}} ch(g)
\end{equation*}
Replacing connections by Dirac operators, the role of the relative Chern--Simons form $cs(d, g^{-1}dg)$ is played by the spectral flow $sf(D, g^{-1}Dg)$ \cite{G93}.

The analytic spectral flow of  a Dirac operator $D$ on  $M$ can be computed as
\begin{equation}
  sf(D, g^{-1}Dg) = \frac{1}{\sqrt{\pi}} \int_0^1 tr(\dot{D}_t e^{-D_t^2}) dt
\end{equation}
where
\begin{equation*}
   D_t = (1-t)D + tg^{-1}Dg, \quad \dot{D_t} = g^{-1}[D, g]
\end{equation*}
The Dirac operator $D$ defines a Fredholm module in K-homology, and the spectral flow computes
the Fredholm index of the Toeplitz operator $PgP$ by the pairing between K-homology and K-theory \cite{G93},
\begin{equation}
   index(PgP)  = \langle [D], [g] \rangle = - sf(D, g^{-1}Dg)
\end{equation}
where $P := (1+D|D|^{-1}) /2$ is the spectral projection.
Baum and Douglas \cite{BD82} first noticed the odd Toeplitz index theorem,
which is an identity connecting the analytical index and topological index,
\begin{equation}
   sf(D, g^{-1}Dg) =   \int_M \hat{A}(M ) \wedge ch(g)
\end{equation}
where $\hat{A}$ is the A-roof genus, for a generalized odd index theorem for manifolds with boundary see \cite{DZ06}.

In particular, we have $\hat{A}(\mathbb{T}^3) = 1$
since $\hat{A}$ is a multiplicative genus and $\hat{A}(S^k) =1$ for spheres.
Hence the degree of $g$ can be computed as the spectral flow on the 3d Brillouin torus,
\begin{equation} \label{odddeg}
   sf(D, g^{-1}Dg) = - \left(\frac{i}{2\pi }\right)^2 \int_{\mathbb{T}^3}  ch(g) = \deg{g}
\end{equation}
Finally, we apply the odd index theorem \eqref{odddeg} to the  Dirac Hamiltonian $H$ of a 3d topological insulator and  the transition matrix $w$ \eqref{wmatrix},
then the Chern--Simons invariant is identified with the spectral flow modulo two,
\begin{equation}
 \upsilon \equiv  sf(H, w^{-1}Hw )  \quad \text{mod 2}
\end{equation}

For manifolds with boundary, a well-defined  Dirac operator involves the APS boundary condition, see \cite{APS76}, \cite{KL04}.
The analytic $\eta$-invariant of a Dirac operator $D$ is defined by
\begin{equation} \label{etaform}
   \eta(D) := \frac{1}{\sqrt{\pi}}   \int_0^\infty  tr(D e^{-sD^2})\frac{ds}{\sqrt{s}}
\end{equation}
If $D_P(t)$ is a family of Dirac operators defining a good boundary value problem, $P$ is a spectral projection, then
the spectral flow of $D_P(t)$ can be computed as variations of the eta invariant, see \cite{G93}, \cite{KL04},
\begin{equation*}
   sf(D_P(t))  =   \tilde{\eta}(D_P(t))|_0^1  - \frac{1}{2} \int_0^1  \frac{d}{dt} \eta (D_P(t)) \, dt
\end{equation*}
where $\tilde{\eta}$ is the reduced eta invariant.
As a remark, if a superconnection $\mathbb{A}$ is used in \eqref{etaform} instead of $D$, then it would define the eta form,
which has important applications in the family index theorem and Witten's holonomy theorem \cite{BF86}.

\subsubsection{Green's function}

In this subsection, we define the Chern--Simons invariant of time reversal invariant topological insulators by Green's functions following \cite{WQZ101}.
This approach has the advantage that
such topological invariant is still valid even in interacting or disordered systems.
The idea of applying Green's functions
to the study of domain walls in D-branes or edge states in topological insulators can be traced back to early works by Volovik,
for example see  \cite{V02}.

By evaluating one loop Feynman diagrams, the second Chern class (\ref{c2eq}) in the 4d action $(\ref{4dactioneq})$  is equivalently defined by fermionic propagators \cite{GJK93}, i.e., Green's functions,
\begin{equation}
  c_2 = \frac{\pi^2}{15} \int \frac{d^4k d\omega}{(2\pi)^5} \epsilon^{\mu\nu\rho\sigma\tau} tr[G \partial_\mu G^{-1}
  G \partial_\nu G^{-1}G \partial_\rho G^{-1}G \partial_\sigma G^{-1}G \partial_\tau G^{-1}]
\end{equation}
where the imaginary-time single-particle Green's function of the  Hamiltonian $H$ is defined as
\begin{equation*}
  G(i \omega, \mathbf{k}) := (i \omega - H(\mathbf{k}))^{-1}
\end{equation*}

For 3d time reversal invariant topological insulators, the electro-magnetic polarization $P_3$  is again obtained by dimensional reduction,
\begin{equation}
   P_3 := \frac{\pi}{6} \int_0^1 du \int \frac{d^3k d\omega}{(2\pi)^{4}} \epsilon^{\mu\nu\rho\sigma} tr[G \partial_\mu G^{-1}
  G \partial_\nu G^{-1}G \partial_\rho G^{-1}G \partial_\sigma G^{-1}G \partial_u G^{-1}]
\end{equation}
By a similar argument as the WZW extension problem, which will be explained in the next subsection, $2P_3$ is well defined up to certain integer belonging to the homotopy group
$   \pi_5(GL(n, \mathbb{C})) \simeq \mathbb{Z} $.
In order to get the topological $\mathbb{Z}_2$ invariant in 2d, we need two WZW extension parameters and to reduce one more dimension.
The 2d topological order parameter is defined similarly,
\begin{equation}
  P_2 := \frac{1}{120} \int_{-1}^1 du dv \int \frac{d^2k d\omega}{(2\pi)^{3}} \epsilon^{\mu\nu\rho} tr[G \partial_\mu G^{-1}
  G \partial_\nu G^{-1}G \partial_\rho G^{-1}G \partial_u G^{-1}G \partial_v G^{-1}]
\end{equation}

The same logic shows that $P_2$ induces the topological $\mathbb{Z}_2$ invariant and $P_2 \equiv 1/2$
corresponds to non-trivial topological insulators in 2d.

\subsubsection{Parity anomaly}

The parity anomaly caused by unpaired Majorana
zero modes fits into an $SU(2)$ gauge theory when the time reversal symmetry is translated into a specific gauge transformation as discussed in \S\ref{CWcorr}.
Fukui, Fujiwara and
Hatsugai \cite{FFH08} explained how to understand the $\mathbb{Z}_2$ invariant
using the analogy of the global $SU(2)$ anomaly  \cite{W82}, which reveals the relation between the second Chern number and
the mod 2 3d topological WZW term, as we now discuss.

As discussed previously, see \S\ref{homotopypar}, the 2nd Chern number $c_2$ for a time reversal invariant model is basically determined by the
integral of the 2nd Chern class over $S^4$. If we decompose $S^4$ into two hemispheres,
then $c_2$ can be computed as the winding number over $S^3$,
\begin{equation}
  c_2 = -\frac{1}{8\pi^2}  \int_{S^4} tr (f^2) = \frac{1}{24\pi^2} \int_{S^3} tr(g^{-1}dg)^3
\end{equation}
where $g: S^3 \rightarrow SU(2)$ is a transition function on the overlap of two hemispheres (i.e. $S^3$), for details see  \cite{N03}. The right hand side is the winding number of $g$ around
the 3-sphere, i.e., the topological WZW term,
and $c_2$ can be viewed as the obstruction to the gauge fixing problem.

Taking the phase ambiguity of a Kramers pair $|\Psi\rangle$ into consideration, when writing $A=\langle \Psi|d|\Psi\rangle$, as explained in
\cite{N03},  the topological space where the transition function really is defined on is
$S^3 \times S^1$.  To accommodate the gauge transformation $r$ for the phase change,
one needs an extra factor of $SU(2)$, due to an extra symmetry this action actually lands in $U(1)
\subset SU(2)$. It is natural, however to embed both these factors
into $SU(3)$, see below. Also,  notice that the boundary $\partial (S^3 \times D^2) = S^3 \times S^1$ and following Witten,
the gauge field $r$ can be extended to $D^2$ as follows,

\begin{equation*}
   \tilde{g}(x) = \begin{pmatrix}	                    g(x)&  \\
                        & 1
                  \end{pmatrix} \in SU(3); \quad
    \tilde{r}(\theta, \rho) = \begin{pmatrix}
                                 1 & & \\
                                 & \rho e^{i\theta} & \sqrt{1-\rho^2} \\
                                 & - \sqrt{1-\rho^2} & \rho e^{-i\theta}
                              \end{pmatrix} \in SU(3)
 \end{equation*}
  ($\rho=1$ gives the $U(1)$ embedding for $r$.)
The  transition function is then given by
\begin{equation*}
   \tilde{g}_r: S^3 \times D^2 \rightarrow SU(3); \,\, (x, \theta, \rho) \mapsto \tilde{r}^{-1}(x,\rho) \tilde{g}(x)\tilde{r}(\theta, \rho)
\end{equation*}
and the 5d topological  WZW term $\Gamma(\tilde{g}_r)$  is defined as usual,
\begin{equation*}
   \Gamma({\tilde{g}_r}) = \frac{-i}{240\pi^2} \int_{S^3 \times D^2} tr(\tilde{g}_r^{-1}d\tilde{g}_r)^5
\end{equation*}
which only takes values that are a multiple of $\pi$, i.e.,  $  \Gamma({\tilde{g}_r})  \in \mathbb{Z} \cdot  \pi$.
Actually, after integrating out the extra variables $\theta$ and $\rho$, we recover the second Chern number again,
\begin{equation}\label{3deq5d}
     \Gamma({\tilde{g}_r}) = \frac{1}{24 \pi} \int_{S^3} tr(g^{-1}dg)^3 = \pi c_2
\end{equation}

We stress that the 5d WZW term is only well-defined up to a multiple of $2\pi$ because of the dependence of  embeddings.
In general, the embedding  to $SU(3)$ is not unique.  Another possible way is to start with the mapping
\begin{equation*}
   G: S^3 \times S^1 \rightarrow SU(2); \quad (x, \theta) \mapsto r^{-1}(\theta)g(x)r(\theta)g^{-1}(x)
\end{equation*}
where  $r(\theta)$ is the diagonal matrix $  r(\theta) = diag( e^{-i\theta/2}, e^{i\theta/2})$.
Since  $\pi_1(SU(2)) = 0$ and $G(x, 0) = G(x, 2\pi) = 1$,  the map $G: S^3 \times S^1 \rightarrow SU(2)$ factors through $S^4$ in this case, analogously to \S\ref{homotopypar}.
As a  result, $G$ falls into two classes according to $\pi_4(SU(2))= \mathbb{Z}_2$.
Now extending $G$ as before, we obtain a map $\tilde{G}$ from $D^5$ into $SU(3)$,
\begin{equation*}
   \tilde{G}(x, \theta, \rho) = {\tilde{g}_r}(x, \theta, \rho){\tilde{g}_r}^{-1}(x, 0, \rho)
\end{equation*}

It was proved in \cite{FFH08} that the corresponding WZW term is the same as before
\begin{equation*}
   \Gamma(\tilde{G}) = \Gamma(\tilde{g}_r)
\end{equation*}
If we further choose a different 5-disc $D'^5$ that bounds the same $S^4$, and then we consider the difference
\begin{equation*}
 \Gamma_{D^5} (\tilde{G}) - \Gamma_{D'^5} (\tilde{G}) =  \frac{-i}{240\pi^2} \int_{S^5} tr(\tilde{G}^{-1}d\tilde{G})^5
\end{equation*}
The right hand side is a multiple of $2\pi$ because of $\pi_5(SU(3)) = \mathbb{Z}$.
This explains why we have to modulo $2\pi$ to eliminate the embedding ambiguity.
Hence  the 5d WZW term is well-defined, i.e., independent of the embedding, modulo $2 \pi$,
and by the relation \eqref{3deq5d} the 3d WZW term is then  $\mathbb{Z}_2$-valued,
\begin{equation}\label{5DZ2}
     \Gamma_{S^3}({g})  \equiv 0, \pi \quad \text{mod } \,\, 2 \pi
\end{equation}
In other words, the mod 2 second Chern number computes the topological $\mathbb{Z}_2$ invariant after embedding $SU(2) \times U(1)\subset SU(2)\times SU(2)$ into $SU(3)$.

Summarizing the results: the phase ambiguity  enlarges
 $S^3$ to $S^3 \times S^1$, which can be replaced by $S^4$ when a special boundary condition is satisfied \cite{FFH08}. With the fixed target space $SU(2)$, more room is made in the parameter space and the
 topological invariant originally assumed to lie in
 $\pi_3(SU(2)) \simeq \mathbb{Z}$ is now in $\pi_4(SU(2)) \simeq \mathbb{Z}_2$.
In this language, a strong $\mathbb{Z}_2$ topological insulator is characterized by a transition matrix \eqref{wmatrix} with odd winding number.

\subsection{Summary}

It is shown in \cite{KLW15, WQZ10} that for a 3d topological insulator, the Chern--Simons invariant and the Kane--Mele invariant
are equivalent, i.e., $\upsilon  = \nu $, since they are just
the integral form and the discrete version of the mod 2 degree.
Hence we call it the $\mathbb{Z}_2$ invariant of time reversal invariant topological insulators and view it as a topological index.
Another proof of the equivalence of the Chern--Simons and Kane--Mele invariants
was given in \cite{FM13} based on the group of reduced topological phases, which will be discussed  in the next section.

Earlier, we discussed a mod 2 index theorem, proved in \cite{FF09}, connecting the analytical index $ind_2(D)$ \eqref{Amod2} and
the topological index \eqref{obsz2} for a Dirac operator with time reversal and chiral symmetries.
Viewing this in the more general framework above, the reason why a mod 2 index theorem appears in $\mathbb{Z}_2$ topological insulators is the following. In an odd dimensional
time reversal invariant fermionic system, the parity anomaly  pops up as a global anomaly, which is really difficult to compute in general  \cite{APM85}.
By translating the problem into a gauge theory, the global parity anomaly is equivalent to a gauge anomaly, which can be dealt with locally.
In our case, the parity of the spectral flow of edge states characterizes the global property of the material,
and the gauge theoretic WZW  term provides a practical way to compute the parity anomaly locally. It is well known that
an index theorem can be used to compute a global quantity by a local formula, for example, the odd index theorem
computes the spectral flow by the odd Chern character as discussed in \S\ref{oddCh}. Therefore, the topological $\mathbb{Z}_2$ invariant
can be interpreted as the mod 2 version of the odd index theorem, which computes the global parity anomaly locally.

Let us recap the relations between different variants of the $\mathbb{Z}_2$  invariant discussed above.
Based on the non-abelian bosonization, the $\mathbb{Z}_2$ invariant describes the parity anomaly of Majorana zero modes in a time reversal invariant topological insulator.
On the one hand, the analytical $\mathbb{Z}_2$ index is the parity of the spectral flow of edge states under an adiabatic evolution. The fermionic path
integral of Majorana zero modes such as  Dirac cones introduces the Pfaffian, while the path integral of the equivalent composite boson
gives the square root of the determinant. After an adiabatic procedure, one can have a different way to bound two Majorana zero modes, so that
the ratio of the effective action, i.e., the sign of the Pfaffian, would change accordingly. The Kane--Mele invariant keeps track of the
mod 2 version of the change of the signs of Pfaffians, that is, the parity of the spectral flow of chiral edge states through the adiabatic evolution.
For infinite dimensional Hilbert spaces, instead of the determinant (resp. Pfaffian) of matrices,
we  consider the determinant (resp. Pfaffian)  line bundles of Dirac operators \cite{F87}.
In the language of bundles, the $\mathbb{Z}_2$ invariant is obtained by
 comparing the orientations of the Pfaffian and the determinant line bundles over the fixed points of the time reversal symmetry.
If we further consider the holonomy of the Pfaffian or the determinant line bundle, the quotient form of the Kane--Mele invariant is  replaced
by a difference form,
since the holonomy of the Pfaffian  or determinant line bundle is an exponentiated eta invariant with a multiple of $i \pi$ \cite{BF86}.
If we model the edge states by Lagrangian sub-manifolds in the symplectic setting, then the spectral flow can be computed by the Maslov
index, that is, the Lagrangian intersection number of the chiral edge states with the Fermi level.

On the other hand, the topological $\mathbb{Z}_2$ index is the topological Wess--Zumino--Witten (WZW) term modulo two,
since the bosonization process delivers the WZW term
as the action functional of chiral edge currents \cite{W84}. The bulk-edge correspondence in this context is given by the Chern--Simons/WZW duality, and
the WZW term is just the topological index of odd Chern characters of gauge transformations.  In our case, the time reversal symmetry induces a specific gauge transform, so
the $\mathbb{Z}_2$ invariant is the mod 2 index of the odd Chern character of this special unitary in the odd K-group.
Hence the mod 2 version of the odd index theorem  computes the spectral flow
of the edge Dirac Hamiltonian, so the topological  $\mathbb{Z}_2$ invariant is interpreted as the Witten index $Tr(-1)^F$.
If we use the Feynman propagator instead of gauge transformations, the WZW term
can be calculated based on the method of Green's functions.
The homotopy theory of Dirac Hamiltonians tells us that the 2nd Chern number, i.e., the integral of the 2nd Chern character
is the fundamental invariant characterizing time reversal invariant models. In addition, the homotopy theory on the effective Hamiltonians
gives us an interesting explanation of the $\mathbb{Z}_2$ invariant.
According to the dimensional ladder in anomalies \cite{N03}, the 3d parity anomaly is the descendant of the 4d chiral anomaly,
so the WZW term (its mod 2 version is the parity anomaly) is naturally connected to
the 2nd Chern character (which computes the chiral anomaly). In other words,
the topological $\mathbb{Z}_2$ invariant can be derived from the 2nd Chern number from the relation between anomalies.
Finally an analogy of the global $SU(2)$ anomaly of the WZW model explains that it is naturally $\Z_2$-valued,
 so we identify the topological $\mathbb{Z}_2$ index as the mod 2 WZW term, and  it is called the parity anomaly instead of the $SU(2)$ anomaly in condensed matter physics.

\section{K-theoretic classification} \label{KThry}

In this section, we will  discuss   the K-theoretic classification of non-interacting (or weak interacting)
Dirac Hamiltonians of topological insulators.
More precisely, the $\mathbb{Z}_2$-valued topological  invariants of topological insulators and Bogoliubov-de Gennes (BdG) superconductors
fit into a periodic table resembling the Bott periodicity as in topological K-theory. First, a Clifford algebra classification for these was
established in  \cite{SRFL09} and the K-theory classification followed as the Clifford extension problem \cite{K09}.
A finer classification of topological
phases based on KR-theory and twisted equivariant K-theory was proposed in \cite{FM13}.
For time reversal invariant topological insulators, KH-theory was also applied to the classification problem in \cite{DG14}.

KR-theory, introduced   by Atiyah \cite{A66}, is the K-theory of Real vector bundles over an involutive space. By contrast,
KH-theory is the K-theory of Quaternionic vector bundles over an involutive space. They are isomorphic to each other by a Fourier transform.
A detailed treatment of twisted equivariant K-theory can be found for example in \cite{FHT11}.

\subsection{Tenfold way}
\label{tenpar}
Generalizing the standard Wigner-Dyson threefold way, namely, symmetries in quantum mechanics described by unitary, orthogonal and symplectic groups,
Altland and Zirnbauer \cite{AZ97} further proposed a tenfold way in random matrix theory, which implies that free fermionic
systems can be classified by ten symmetry classes. Indeed, there are three types of discrete (pseudo)symmetries in topological insulators and
BdG Hamiltonians:
the time reversal symmetry $\mathcal{T}$, the particle-hole symmetry $\mathcal{P}$ and the chiral symmetry $\mathcal{C}$,
and the combinations of these three symmetries give ten  classes in total.

Based on a time reversal invariant Dirac Hamiltonian $H$, which means that $\mathcal{T} H \mathcal{T}^{-1} = H$,  and additionally $\mathcal{T}^2 =\pm 1$ depending on the spin being integer or half-integer,
 the time reversal symmetry (TRS) induces three classes
\begin{equation}
    TRS = \left\{
  \begin{array}{l l}
    +1 & \quad \text{if} \quad \mathcal{T} H(\mathbf{k}) \mathcal{T}^{-1} = H(- \mathbf{k}), \,\, \mathcal{T}^2=+1\\
    -1 & \quad \text{if} \quad \mathcal{T} H (\mathbf{k})\mathcal{T}^{-1} = H(-\mathbf{k}), \,\, \mathcal{T}^2 =-1 \\
    0  & \quad \text{if} \quad \mathcal{T} H(\mathbf{k}) \mathcal{T}^{-1} \neq H(-\mathbf{k})
  \end{array}  \right.
\end{equation}
Similarly, the particle hole symmetry (PHS) also gives three classes,
\begin{equation}
    PHS = \left\{
  \begin{array}{l l}
    +1 & \quad \text{if} \quad \mathcal{P} H(\mathbf{k}) \mathcal{P}^{-1} =- H(\mathbf{k}), \,\, \mathcal{P}^2=+1\\
    -1 & \quad \text{if} \quad \mathcal{P} H(\mathbf{k}) \mathcal{P}^{-1} = -H(\mathbf{k}), \,\, \mathcal{P}^2 =-1 \\
    0  & \quad \text{if} \quad \mathcal{P} H(\mathbf{k}) \mathcal{P}^{-1} \neq -H(\mathbf{k})
  \end{array}  \right.
\end{equation}
The chiral symmetry can be defined by the product $\mathcal{C} = \mathcal{T} \cdot \mathcal{P}$, sometimes also referred to as the sublattice symmetry.
Since $\mathcal{T}$ and $\mathcal{P}$ are anti-unitary, $\mathcal{C}$ is a unitary operator.
If both $\mathcal{T}$ and $\mathcal{P}$ are non-zero, then the chiral symmetry is present, i.e., $\mathcal{C} = 1$.
On the other hand, if both $\mathcal{T}$ and $\mathcal{P}$ are zero,
then $\mathcal{C}$ is allowed to be either $0$ (type A or unitary class, according to the classification of \cite{AZ97}, see e.g.\ \cite{SRFL09}) or $1$ (type AIII or chiral unitary class).
In sum, there are $3 \times 3 +1 =10$ symmetry classes. In particular,
the half-spin Hamiltonian with time reversal symmetry falls into  type AII or symplectic class,
which is the case we are mostly interested in, for more details about the 10-fold way and other symmetry classes see \cite{SRFL09}.

In \cite{K09}, the above particle-hole symmetry $\mathcal{P}$ is replaced by a $U(1)$-symmetry  $\mathcal{Q}$ representing the charge or particle number.
Roughly, the parity $\mathcal{P}$ is related to the charge by $\mathcal{P} = (-1)^\mathcal{Q} $.
So sometimes topological insulators are also referred to as $U(1)$ and time reversal $\mathbb{Z}_2$ symmetry protected phases.

\subsection{Nonlinear $\sigma$-model}
\label{genhompar}
One classification scheme is given by the homotopy groups of  non-linear sigma models or target spaces of the Brillouin torus \cite{SRFL09}.
Assume that there are $N$ occupied bulk bands and denote the occupied bands by $| \mathbf{k}\rangle: = \sum_{n=1}^N | n, \mathbf{k} \rangle$.
Further assume that there are also $N$ unoccupied bands and that there exists a band gap between the conduction (occupied) and valence (unoccupied) bands, and the Fermi level $E_F$ is assumed to lie in the band gap. If there is a physical material the Fermi level is determined by it, purely mathematically, without a given material, one can choose $E_F$
to be any energy in the band gap.
Define the Fermi projection in ket-bra notations as
\begin{equation}
   P_F := \sum_{n=1}^N  | n, \mathbf{k}\rangle \langle n, \mathbf{k}| = | \mathbf{k}\rangle \langle \mathbf{k}|
\end{equation}
In addition, the Fermi level is always assumed to be zero, i.e., $E_F = 0$, so that the occupied bands have negative eigenvalues.
 From now on we denote the Fermi projection by $P \equiv P_F$ for simplicity.

For non-degenerate band structures, the spectral flattening trick can always be applied,
which replaces the $i$-th band  $E_i(\mathbf{k})$ (a continuous function over the Brillouin torus) by a constant function $E_i(\mathbf{k}) = E_i$.
Define the sign of the Dirac Hamiltonian  by $F(\mathbf{k}) := H(\mathbf{k})|H(\mathbf{k})|^{-1}$, by which the band structure is reduced to
a flattened two-band model, or in terms of the Fermi projection the flattened Hamiltonian $F$ is written as
\begin{equation*}
   F(\mathbf{k}) = 1-2P
\end{equation*}
We use the notation $F$ to indicate that it looks like the grading operator of a Fredholm module in K-homology satisfying
\begin{equation*}
  F^2 = 1, \quad F^* = F
\end{equation*}
For the unitary class (type A), the flattened Hamiltonian $F(\mathbf{k})$ defines a  map from the Brillouin torus
to the Grassmannian, aka.\ a non-linear sigma model:
\begin{equation}
   F: \mathbb{T}^d \rightarrow Gr(N, \mathbb{C}^{2N}) = U(2N) / (U(N) \times U(N))
\end{equation}
since the occupied and empty bands each have $N$ levels by assumption and we don't distinguish energy levels among occupied or empty bands.
Similarly, for the symplectic class (type AII),
$F$ defines another non-linear sigma model,
\begin{equation}
      F: \mathbb{T}^d \rightarrow Gr(N, \mathbb{R}^{2N}) =O(2N) / (O(N) \times O(N))
\end{equation}

Without loss of generality, we assume that the empty bands have infinitely many levels. In other words,
the $2N$-dimensional total  vector space  is replaced by the infinite dimensional Hilbert space,
then the classifying space $BU(N)$ (resp. $BO(N)$) plays the role of the Grassmannian $Gr(N, \mathbb{C}^{2N})$ (resp. $Gr(N, \mathbb{R}^{2N})$).
If $N$ is big enough, the lower dimensional homotopy groups stabilize.
Accordingly one can regard the target spaces the direct limits,
\begin{equation*}
   BU = \varinjlim_{N}BU(N), \quad  BO = \varinjlim_{N}BO(N)
\end{equation*}
which capture the stable (under increasing $N$) features, which is what stable homotopy theory is concerned with.

For a compact Hausdorff space $X$, the topological K-theory $K(X)$ is defined to be the Grothendieck group of stable isomorphism classes of
finite dimensional vector bundles over $X$.
The complex or real K-group can be equivalently defined by homotopy classes of maps from $X$ to the classifying spaces,
\begin{equation*}
   K(X) = K_\mathbb{C}(X) \simeq [X, BU \times \mathbb{Z}], \quad KO(X) = K_\mathbb{R}(X) \simeq [X, BO\times \mathbb{Z}]
\end{equation*}
In stable homotopy theory, we say the spectrum of the complex K-theory is $  BU \times \mathbb{Z}$, which has a period 2, i.e.,
$    \Omega^2BU \simeq BU  \times \mathbb{Z}   $,
where $\Omega $ is the loop space operation.
Similarly, the spectrum of the real K-theory is $  BO \times \mathbb{Z}$, which has a period 8, i.e.,
$   \Omega^8BO  \simeq  \mathbb{Z} \times BO  $.
Such periodic phenomena, are known as the Bott periodicity. The first case was discovered by Bott
in the study of homotopy groups of spheres using Morse theory.
For unitary or orthogonal groups,  Bott periodicity is spelled out as
\begin{equation*}
  \pi_{k+1}(BU) =  \pi_{k}(U) \simeq \pi_{k+2}(U), \quad \pi_{k+1}(BO) = \pi_k(O) \simeq \pi_{k+8}(O)
\end{equation*}
Besides in topological K-theory and homotopy groups of classical groups,   Bott periodicity is also present
in Clifford algebras and their representations. Of course,  Clifford modules and topological K-theory are intimately related in index theory.

Going back to band theory, in the language of K-theory via stabilizing (embedding into the direct limit) the non-linear sigma model induces an element in the complex or real K-group, the homotopy class of the map $\TT^d\stackrel{F}{\to} Gr(N)\to BU(N)\to BU$.
\begin{equation}
 [F] \in \left\{
  \begin{array}{l l}
    K(\mathbb{T}^d ) & \quad \text{for type A} \\
    KO(\mathbb{T}^d) & \quad \text{for type AII}
  \end{array}  \right.
\end{equation}
Furthermore, if the map from the Brillouin torus can transformed into  a map $\hat F$ from the sphere $S^n$ either by a quotient map or by extending and quotienting, as described in the previous section, then
\begin{equation}
   [\hat F] \in \left\{
  \begin{array}{l l}
    \pi_n (BU) = \pi_{n-1} (U)& \quad \text{for type A} \\
    \pi_n(BO)=\pi_{n-1}(O) & \quad \text{for type AII}
  \end{array}  \right.
\end{equation}
For instance, consider the edge Hamiltonian in the quantum spin Hall effect, i.e., $n =1$, then the 2d $\mathbb{Z}_2$ invariant
corresponds to
\begin{equation*}
     \pi_1(BO)=\pi_{0}(O) \simeq \mathbb{Z}_2
\end{equation*}

\subsection{Clifford modules}
Following \cite{K09}, we briefly go over the classification approach via the Clifford extension problem in this subsection.
For this one regards free fermions in the basis of  operators
$\hat{c}_n$ which satisfy the Clifford relations $\hat{c}_l\hat{c}_m+\hat{c}_m\hat{c}_l=2\delta_{l,m}$.

Then  the $U(1)$ symmetry $\mathcal{Q}$ and conjugation by the time reversal symmetry $\mathcal{T}$ have  matrix representations on the $\hat{c}_j$
\begin{equation}
Q = i\sigma_0 \otimes \sigma_y, \quad T = -i\sigma_y \otimes \sigma_z
\end{equation}
where $\sigma_0 = \begin{pmatrix}
                   1 & 0 \\
                   0 & 1
                  \end{pmatrix}
$ and $\sigma_i$ are the Pauli matrices.
\begin{equation*}
  Q^2= T^2 = -1, \quad \{Q, T \} = 0
\end{equation*}
A Dirac Hamiltonian $H=\frac{i}{4}\sum A_{jk}\hat{c}_j\hat {c}_k$ represented by  a real skew-symmetric 2N$\times$2N matrix $A$
has  $U(1)$ symmetry if and only if $[Q, A] = 0$. Similarly, $H$ being time reversal invariant is equivalent to
$\{T, A \} =0$. To simplify things, one can rescale $A$, e.g.\ by regarding it as a skew--symmetric quadratic form or $iA$ as a Hermitian form,
to have Eigenvalues $i,-i$. Denote the result by $\tilde A$, then
   $\tilde{A}^2 = -1$

Finding a compatible matrix $A$  reduces to the problem of finding an extension of a Clifford algebra with $n$ generators by one more generator $\tilde A$.
Here $n=0$ for no symmetry, $n=1$ for $\mathcal{T}$ only and $n=2$ for $\mathcal{T}$ and $\mathcal{Q}$. For example for $n=2$:
 $e_1 = \mathcal{T}$ and $e_2 = \mathcal{QT}$ and we want to add one more generator $e_3 = \tilde{A}$ for the
Dirac Hamiltonian with both time-reversal symmetry and $U(1)$ symmetry.
In general, the Clifford extension problem is given by  considering the extensions
\begin{equation*}
  i : Cl_{0, n}(\mathbb{R}) \hookrightarrow Cl_{0, n+1}(\mathbb{R})
\end{equation*}
Here $Cl_{p,q}$ is the real Clifford algebra for the quadratic form with signature $(\underbrace{+,\dots,+}_{p},\underbrace{-,\dots,-}_{q})$.

 Bott periodicity in Clifford algebra manifests itself as Morita equivalence,
\begin{equation*}
   \begin{array}{ll}
         Cl_{n+2}(\mathbb{C}) \simeq M_2(Cl_n(\mathbb{C})) \sim_M Cl_n(\mathbb{C}), \\
        Cl_{p, q+8}(\mathbb{R}) \simeq Cl_{p+8, q}(\mathbb{R}) \simeq M_{16}(Cl_{p,q}(\mathbb{R})) \sim_M Cl_{p,q}(\mathbb{R})
   \end{array}
\end{equation*}
where $M_n(R)$ is the $n \times n$ matrix of some commutative ring $R$ and $\sim_M$ stands for the Morita equivalence in ring theory.
Recall that two unital rings are Morita equivalent if they have equivalent categories of left modules.

In order to classify the Dirac Hamiltonians with or without the above mentioned $\mathbb{Z}_2$ symmetries,
we actually deal with the representations of the above Clifford algebras, i.e. Clifford modules, since the $e_i$ are represented by matrices.
Denote by $M(Cl_{0,n}(\mathbb{R}))$  the free abelian group
generated by irreducible $\mathbb{Z}_2$-graded $Cl_{0,n}(\mathbb{R})$-modules.
There exists a restriction map induced by  $i$,
\begin{equation*}
  i^* :  M(Cl_{0,n+1}(\mathbb{R})) \rightarrow M(Cl_{0,n}(\mathbb{R}))
\end{equation*}
Each Clifford extension $i$ defines the cokernel of $i^*$,
\begin{equation}
   A_n := coker (i^*) = M(Cl_{0,n}(\mathbb{R}))/ i^*M(Cl_{0,n+1}(\mathbb{R}))
\end{equation}
For the complex case, we define $M^c(Cl_{0,n}(\mathbb{R}))$ as the Clifford modules of the complexification
$Cl_n(\C)=Cl_{0,n}(\mathbb{R}) \otimes_\mathbb{R} \mathbb{C}$, $A_n^c$ can be obtained similarly.
It is $A_n$ (or $A_n^c$) that encodes the information about non-trivial extensions of free Dirac Hamiltonians of different dimensions in the periodic table.

Furthermore, the Clifford modules and the cokernels inherit the Bott periodicity:
\begin{equation*}
   A_{k+8} \simeq A_k, \quad A_{k+2}^c \simeq A_k^c
\end{equation*}
In fact, the cokernels have the same form as the homotopy groups of the classifying spaces \cite{ABS64},
\begin{equation*}
   A_{n} \simeq \pi_{n}(BO), \quad A_{n}^c \simeq \pi_{n}(BU)
\end{equation*}
For instance, if we extend  $Cl_{0,1}(\mathbb{R})$ generated by $\mathcal{T}$ in the quantum spin Hall effect,
\begin{equation*}
   i: Cl_{0,1}(\mathbb{R}) \simeq \mathbb{C} \rightarrow  Cl_{0,2}(\mathbb{R}) \simeq \mathbb{H}
\end{equation*}
then $A_1 \simeq \mathbb{Z}_2$ gives the 2d $\mathbb{Z}_2$ invariant. One step further, if we extend $Cl_{0,2}$
generated by $\mathcal{T}$ and $\mathcal{QT}$ for 3d topological insulators,
\begin{equation*}
   i: Cl_{0,2}(\mathbb{R}) \simeq \mathbb{H} \rightarrow  Cl_{0,3}(\mathbb{R}) \simeq \mathbb{H} \oplus  \mathbb{H}
\end{equation*}
then $A_2  \simeq \mathbb{Z}_2$ shows that there exists another $\mathbb{Z}_2$ invariant in 3d.

It is easy to promote vector spaces to vector bundles and get the K-theory of Clifford bundles,
that is, to pass from the Clifford algebra classification to the K-theory classification of topological insulators.
For any pseudo--Riemannian vector bundle $E \rightarrow X$ over a compact Hausdorff space $X$, define an associated  Clifford bundle
$Cl(E) \rightarrow X$ as usual, and a representation gives a $\mathbb{Z}_2$-graded vector bundle $M(E) \rightarrow X$.
The Clifford extension problem arises naturally when we add a one-dimensional trivial bundle $T_1 \rightarrow X$ by the Whitney sum, and the cokernel
bundle is defined as
\begin{equation*}
  A(E) := coker (i^* : M(E \oplus T_1) \rightarrow M(E))
\end{equation*}

By the Clifford grading $M(E) = M^0 \oplus M^1$ and its Euler class is defined as the formal difference
\begin{equation*}
   \chi_E(M) := [M^0] - [M^1]
\end{equation*}
which induces a map from the cokernel bundle to the reduced KO-group,
\begin{equation*}
   \chi_E: A(E) \rightarrow KO( B(E), S(E)) = \widetilde{KO}(E^+)
\end{equation*}
where the right hand side is the Thom construction, $B(E)$ and $S(E) $ are the ball and sphere bundles of $E$, $E^+$ is the  one-point compactification.
In particular, if the vector bundle $E$ is of rank $n$, and assume the base space is contractible $X \sim pt$, then $A(E)$ is reduced to $A_n$
and we get the isomorphism
\begin{equation*}
   A_n \simeq  KO(S^n) =\pi_{n}(BO)
\end{equation*}
More details can be found in \cite{ABS64}.

On the level of Clifford modules, the Clifford extension problem is described by the triple $(E, F; \sigma)$, where
$E, F$ are elements in  $M(Cl_{0, n+1}(\mathbb{R}))$ and $\sigma$ is a linear orthogonal map identifying $E|_{Cl_{0,n}} = F|_{Cl_{0,n}}$
as restricted representations of $Cl_{0, n}(\mathbb{R})$, which has a direct generalization into Clifford bundles.
In general, for a pair of spaces $(X, Y)$, given two vector bundles $E, F$ on $X$ and a bundle map $\sigma : E \rightarrow F$
such that $\sigma$ is an isomorphism $E|_Y \simeq F|_Y$ when restricted to $Y$,
the Atiyah-Bott-Shapiro construction \cite{ABS64} defines an element $[(E, F; \sigma)]$ in the relative K-group $K(X, Y) : = \tilde{K}(X/Y)$.
In \cite{K09}, Kitaev applied the  Atiyah-Bott-Shapiro construction to the Clifford extension problem on Clifford bundles and proposed the K-theory classification scheme for topological insulators.

\subsection{Twisted equivariant K-theory}
\label{Ktheorypar}

An even more general framework has been set up in \cite{FM13} for studying  topological phases of quantum systems with additional symmetries
by twisted equivariant K-theory.
For a symmetry group $G$, a homomorphism $\phi: G \rightarrow \{ \pm 1 \}$ is defined to keep track of the unitary or anti-unitary symmetry,
and a group extension $\tau$,
\begin{equation*}
   1 \rightarrow \mathbb{T} \rightarrow G^{\tau} \rightarrow G \rightarrow 1
\end{equation*}
is introduced to take care of the phase ambiguity in quantum mechanics.
As a topological invariant, the group of reduced topological phases  $RTP(G, \phi, \tau)$ is defined to be the abelian group
of generalized quantum symmetries $(G, \phi, \tau)$. In this subsection, we will see how to apply twisted equivariant K-theory to classify
topological phases closely following \cite{FM13}.

For time reversal invariant topological insulators, the reduced topological phase group can be computed by the shifted KR-group of the Brillouin torus,
\begin{equation}
   RTP(G, \phi, \tau) \simeq KR^{-4}(\mathbb{T}^d)
\end{equation}
Similarly, if the parity is reversed (i.e., with particle-hole symmetry $\mathcal{P}$ and $\mathcal{P}^2 =-1$), it can be computed by the equivariant K-group,
\begin{equation}
      RTP(G, \phi, \tau) \simeq K_{\mathbb{Z}_2}(\mathbb{T}^d)
\end{equation}
Finally, if $\mathcal{T}^2= -1$ and $\mathcal{P}^2 = -1$, then the twisted equivariant K-groups are applied,
\begin{equation}
   RTP(G, \phi, \tau) \simeq KR^{-4}_{\mathbb{Z}_2} (\mathbb{T}^d)  \simeq KO^{-4}_{\mathbb{Z}_2}(\mathbb{T}^d)
\end{equation}

For the  torus $\mathbb{T}^d$, the KR-groups can be computed by the stable splitting of the
torus into wedge products of spheres and the twisted Thom isomorphism.
Using the reduced KR-theory defined as usual $   KR(X) \simeq \widetilde{KR}(X) \oplus \mathbb{Z}$, we have
\begin{equation*}
   \widetilde{KR}^{-4} (\mathbb{T}^2) \simeq  \mathbb{Z}_2, \quad  \widetilde{KR}^{-4} (\mathbb{T}^3)  \simeq (\mathbb{Z}_2)^{\times 4}
\end{equation*}

Let us look closely at two natural maps appeared in the computation of $\widetilde{KR}^{-4}(\mathbb{T}^3)$.
One is the projection map, i.e., three projections $   p_{ij}: \mathbb{T}^3 \rightarrow \mathbb{T}^2 $,
which induce three embeddings,
\begin{equation*}
   p_{ij}^*:  \widetilde{KR}^{-4} (\mathbb{T}^2) \rightarrow  \widetilde{KR}^{-4} (\mathbb{T}^3), \quad i.e. \quad  \mathbb{Z}_2 \hookrightarrow (\mathbb{Z}_2)^{\times 4}
\end{equation*}
The other one is the collapse map $  q: \mathbb{T}^3 \rightarrow S^3 $,
which also induces a map between the KR-groups,
\begin{equation*}
   q^* :  \widetilde{KR}^{-4} (S^3) \rightarrow  \widetilde{KR}^{-4} (\mathbb{T}^3)
\end{equation*}
Further, it was shown in \cite{FM13} that its image gives the fourth $\mathbb{Z}_2$ component in $\widetilde{KR}^{-4} (\mathbb{T}^3)$,
\begin{equation*}
  Im(q^*) \simeq \mathbb{Z}_2  \hookrightarrow  (\mathbb{Z}_2)^{\times 4}
\end{equation*}
which corresponds to the $\mathbb{Z}_2$ invariant for strong  topological insulators. First note that the stable splitting of $\mathbb{T}^3$ is given by
\begin{equation*}
  \mathbb{T}^3 \sim_{stable} S^1 \vee S^1 \vee S^1 \vee S^2 \vee S^2 \vee S^2\vee S^3
\end{equation*}
The idea behind the proof is that
a canonical element in $\widetilde{KR}^{-4} (S^3)$ is obtained by constructing a principal
$Sp(1)$-bundle $E$ on $S^3$, then pull it back  to get an element   $q^*E \in \widetilde{KR}^{-4} (\mathbb{T}^3)$.
Taking the time reversal symmetry into account, the image  $q^*E$ is then identified  with the fourth component
$\mathbb{Z}_2 \in \widetilde{KR}^{-4} (S^3)$ from the stable splitting. In the 2d case the situation is similar and actually one can prove which component is used by looking carefully at the construction of \S\ref{homotopypar}.

The group of reduced topological phases is important in that it provides a universal topological invariant.
Recall that $\nu$ is the Kane--Mele invariant defined by the change of the signs of Pfaffians, and
$\upsilon$ is the Chern-Simons  invariant defined by  the mod 2 WZW term.
For 3d topological insulators, Freed and Moore \cite{FM13} gave another proof of the equivalence
$\nu =\upsilon$ by showing that both $\nu$ and $\upsilon$ factor through the abelian group $RTP(G, \phi, \tau) = {KR}^{-4} (\mathbb{T}^3)$,
furthermore both vanish on $Im(p_{ij}^*)$ and equal  the identity on $Im(q^*)$.

\section{Further Discussions}

\subsection{Bulk-edge correspondence revisited}
In this subsection, let us go over the bulk-edge correspondence in topological insulators again on different levels.
We will relate it to a generalized index theorem for further investigation.
In general, the bulk-edge correspondence is a realization of the holographic principle in condensed matter physics.
 The philosophy of the holographic principle is that the boundary field theory determines the geometry in the bulk, and
conversely the bulk geometry determines the boundary theory up to some equivalence.
Even though the general holographic principle  still remains a conjecture, the Chern--Simons/Wess--Zumino--Witten (CS/WZW) correspondence
is well understood as an actual theorem, which actually gives the bulk-edge correspondence for 3d topological insulators on the action functional level.

On the level of effective Hamiltonian, the 1d edge Hamiltonian is obtained from the 2d bulk Hamiltonian by a partial Fourier transform along the
direction with translational invariance in \S \ref{BEcorr}. Such correspondence realized by the partial Fourier transform  holds for any dimension,
so there exists a one-to-one correspondence between the bulk and edge Hamiltonians.

For 3d topological insulators, the $\mathbb{Z}_2$ invariant is the mod 2 WZW term, which is derived from the Chern--Simons theory in \S\ref{CWcorr}.
In other words, the bulk-edge correspondence on the level of action functional
is given by the duality between the bulk Chern--Simons action and the boundary topological WZW term.
If we consider quantum field theories, then the bulk is described by a topological quantum field theory (TQFT), and the boundary
is characterized by a rational  conformal field theory (RCFT). So  the bulk-edge correspondence on the level of field theory
is given by CS3/WZW2  as a  correspondence between 3d TQFT and 2d RCFT. On the other hand, based on the gauge group $G$ such as $SU(2)$,
there exists a one-to-one correspondence between the WZW models and Chern--Simons theories, both characterized by
their level $k \in \mathbb{Z}$.
So on the level of group cohomology, the bulk-edge correspondence is given by the Dijkgraaf-Witten theory \cite{DW90}.
If the Dijkgraaf-Witten transgression map is
 $\tau : H^4(BG, \mathbb{Z}) \rightarrow H^3(G, \mathbb{Z})$,
then $\tau$ is the correspondence between the Chern--Simons theory and WZW model associated with $G$,
for the geometric construction on bundle gerbes, see \cite{CJMSW05}.

The generalized index theorem in \S\ref{BEcorr} can be viewed as a  bulk-edge correspondence on the level of index theory. With the presence of
a domain wall, two quantum Hall states characterized by distinct first Chern numbers are separated by a boundary
created by the domain wall.  The variation of the first Chern numbers across the domain wall
is identified with the spectral flow of the edge Hamiltonian along the boundary. In this model, the index map of the extended Hamiltonian plays the role of a correspondence
connecting the topological invariants from the bulk and the analytic behavior of the edge states.

For 3d $\mathbb{Z}_2$ topological insulators, we have a similar picture as that in the above quantum Hall model. With the presence of the time reversal symmetry,
two 3d bulk systems characterized by distinct Chern-Simons theories are separated by a boundary consisting of the fixed points of the time reversal symmetry.
If the time reversal symmetry is encoded into a specific gauge transformation, then the variation of the bulk Chern-Simons actions is exactly the topological
WZW term of this specific gauge transformation. By the global $SU(2)$-anomaly, the WZW term is indeed $\mathbb{Z}_2$-valued. On the other side,
the Pfaffian formalism of the Kane--Mele invariant
counts the parity of the spectral flow of the chiral edge state along the boundary, i.e., the fixed points. Now the correspondence, i.e., the
index map, is given by the Witten index $Tr(-1)^F$ since it counts the parity of Majorana zero modes such as Dirac cones.
Hence the story of the topological $\mathbb{Z}_2$ invariant can be interpreted as a bulk-edge correspondence realized by the Witten index.

If we model the bulk and boundary spaces by $C^*$-algebras, then the bulk-edge correspondence can be realized by Kasparov module in bivariant K-theory.
So on the level of Kasparov's analytic bivariant K-theory, the bulk-edge correspondence is the Kasparov product with a carefully selected Kasparov module according to the interpretation of index maps in KK-theory.
For quantum Hall effect, such a bulk-edge correspondence is explicitly constructed in \cite{BCR14}, where the authors
represent the bulk conductance as a Kasparov product of the boundary K-homology with a specific extension class (i.e. $KK^1$-class) as the correspondence.
A similar model for time reversal invariant topological insulators
is expected to exist, and the bulk-edge correspondence is expected to be realized  by Kasparov product in real KK-theory.

\subsection{Noncommutative geometry}
There are different sources of noncommutativity in condensed matter physics, which call for generalizations of classical models
in the noncommutative world.
For instance, the introduction of magnetic fields into a lattice model breaks the  translational symmetry, that is, the effective Hamiltonian
does not commute with the translation operators any longer, which is the key assumption for the classical Bloch theory of crystal wavefunctions.
Another source of noncommutativity is created by nontrivial quasi-particle statistics such as non-abelian anyons in fractional quantum Hall systems.
In addition, systems with impurities or defects such as  disordered topological insulators are of realistic importance, which is the main source of noncommutative geometry models up to date.

Noncommutative geometry \cite{C94} has already achieved some success in modeling the integer quantum Hall effect and topological insulators with disorders.
The work  by Bellissard, van Elst and Schulz-Baldes \cite{BES94} set up the framework for studying topological invariants of a disordered system
by noncommutative geometry. The disordered integer quantum Hall effect can be characterized by
the  noncommutative first Chern number, which  was obtained  by combining the Connes--Chern character with the Kubo--Chern formula \cite{BES94}.
The  noncommutative 2nd Chern number is a natural generalization for time reversal invariant models, and  the general  noncommutative n-th Chern number was also studied in \cite{PLB13}.

A survey on the  noncommutative approach to disordered topological insulators was written by Prodan \cite{P11}.
The   noncommutative odd Chern character was discussed in a recent paper \cite{PS14} by Prodan and Schulz-Baldes
in the same framework as in \cite{BES94}. A noncommutative formula for the isotropic magneto-electric response of
disordered insulators under magnetic fields was also proposed in \cite{LP13}.
It would be interesting to see more noncommutative generalizations of both the analytical and topological $\mathbb{Z}_2$ index
of topological insulators in the future.

\section{Acknowledgments}
The authors would like to thank the Max-Planck Institute for Mathematics in Bonn for its hospitality. Major parts of this paper were written there.
We would like  thank Charles Kane and Jonathan Rosenberg for discussions and inspiration. We would also like to thank
  Barry Simon, Joel Moore, Duncan Haldane  and Masoud Khalkhali, in chronological order, for further useful discussions at the different stages of the project.
RK  thankfully  acknowledges  support  from  the  Simons foundation under
collaboration grant \# 317149 and BK thankfully  acknowledges  support  from  the NSF
under the grants PHY-0969689 and PHY-1255409.

%\newpage

\nocite{*}
\bibliographystyle{plain}
\bibliography{SurveyTI}

\end{document}